\documentclass{article}

\usepackage{arxiv}

\usepackage[utf8]{inputenc} 
\usepackage[T1]{fontenc}    
\usepackage{hyperref}       
\usepackage{url}            
\usepackage{booktabs}       
\usepackage{amsfonts}       
\usepackage{nicefrac}       
\usepackage{microtype}      
\usepackage{lipsum}
\usepackage{graphicx}
\usepackage{siunitx}
\usepackage{amsmath}
\usepackage{authblk}
\usepackage{accents}
\usepackage{pdfprivacy}

\newcommand{\bu}[1]{\mathbf{#1}}

\newcommand{\intt}{\int\limits_{a}^{b}}
\newcommand{\lt}{$a$}
\newcommand{\ut}{$b$}
\newcommand{\ltt}{a}
\newcommand{\utt}{b}

\newcommand{\lY}{\underbar{$Y$}}
\newcommand{\uY}{\overline{Y}}

\newcommand{\Ftt}{F(\theta)}
\newcommand{\ftt}{f(\theta)}

\newcommand{\Ft}{$F$(\theta)}

\newcommand{\lF}{\underbar{$F$}(\theta)}
\newcommand{\lFp}{\underbar{$F$}(p)}

\newcommand{\lFF}{\underbar{F}(\theta)}
\newcommand{\lFFD}[1]{\underbar{F}_{\mathcal{D}_{#1}}(\theta)}

\newcommand{\uF}{\overline{F}(\theta)}
\newcommand{\uFp}{\overline{F}(p)}

\newcommand{\uFFD}[1]{\overline{F}_{\mathcal{D}_{#1}}(\theta)}
\newcommand{\lG}{\underbar{$G$}(\theta)}
\newcommand{\uG}{\overline{G}(\theta)}

\newcommand{\mc}[1]{\mathcal{#1}}

\DeclareMathOperator*{\argmax}{arg\,max}

\title{Probability bound analysis: A novel approach for quantifying parameter uncertainty in decision-analytic modeling and cost-effectiveness analysis}

\author[1,2]{Rowan Iskandar}

\affil[1]{Center of Excellence in Decision-Analytic Modeling and Health Economics Research, Swiss Institute for Translational and Entrepreneurial Medicine, sitem-insel AG, Bern, 3010, Switzerland}
\affil[2]{Department of Health Services, Policy, \& Practice, Brown University, Providence, RI, 02906, USA}

\begin{document}
\maketitle
\begin{abstract}
Decisions about health interventions are often made using limited evidence. Mathematical models used to inform such decisions often include uncertainty analysis to account for the effect of uncertainty in the current evidence base on decision-relevant quantities. 
However, current uncertainty quantification methodologies, including probabilistic sensitivity analysis (PSA), require modelers to specify a precise probability distribution to represent the uncertainty of a model parameter. 
This study introduces a novel approach for propagating parameter uncertainty, probability bounds analysis (PBA), where the uncertainty about the unknown probability distribution of a model parameter is expressed in terms of an interval bounded by lower and upper bounds on the unknown cumulative distribution function (p-box) and without assuming a particular form of the distribution function. 
We give the formulas of the p-boxes for common situations (given combinations of data on minimum, maximum, median, mean, or standard deviation), describe an approach to propagate p-boxes into a black-box mathematical model, and introduce an approach for decision-making based on the results of PBA. 
We demonstrate the characteristics and utility of PBA versus PSA using two case studies. 
In sum, this study provides modelers with practical tools to conduct parameter uncertainty quantification given the constraints of available data and with the fewest assumptions.
\end{abstract}
\section{Introduction}
Decision-analytic models (DAMs) have been used in numerous applications, from clinical decision-making to cost-effectiveness analysis (CEA). 
A DAM integrates evidence within a coherent and explicit mathematical structure used to link evidence to decision-relevant outcomes.\cite{hunink2014decision}
There are situations where the evidence required for informing the values of model parameters that govern the behavior of DAMs is incomplete or non-existence, for example, health economic modeling at the early stages of a product's life cycle \cite{drummond2020modeling,rothery2017characterising} and lack of resources to obtain the required data.\cite{world2019guide}
The importance of explicitly accounting for incomplete knowledge about model parameters (\textit{parameter uncertainty}) and propagating its effect through a decisional process is underscored in numerous guidance documents in health, including, but not limited to, the guidelines by the International Society for Pharmacoeconomics and Outcomes Research (ISPOR)-Society for Medical Decision Making (SMDM),\cite{briggs2012model} the Agency for Healthcare Research and Quality (AHRQ),\cite{dahabreh2016recommendations} the 2nd panel for Cost-Effectiveness Analysis (CEA) in Health and Medicine,\cite{sanders2016recommendations} and beyond.\cite{national2012assessing}
At its most basic characterization, parameter uncertainty means that we do not know the exact value of a parameter as several different (potentially uncountable) values may be possible for reasons such as the amount (the size of the available samples of observations) and quality (measurement error or accuracy of the observations) of the available information.\cite{helton2011quantification} 
In many situations, the only information we have about a parameter is that it belongs to an interval bounded by a lower bound and an upper bound.
In addition to knowing the interval, we may have some information about the relative plausibility of different values of $\theta$ in the interval.
In situations where we have data or previous knowledge about a parameter, we can leverage standard statistical techniques to represent uncertainty in the form of a probability distribution.
However, when data and knowledge are limited, we may have only partial or no information about the probability distribution, i.e., we can not assign the relative plausibilities of different parameter values. 
In some cases, we only know the measures of central tendency (mean or median) from published papers, while, in more extreme cases, only the minimum and maximum values are known to the researchers.
To handle such data sparsity situations, it is necessary to have an approach for quantifying parameter uncertainty using the fewest number of assumptions and without the need for assuming precise probability distributions.
\\ \\
Despite the emphasis on its importance, the ISPOR-SMDM best-practice \cite{briggs2012model} recommends only two analytical tools for evaluating the effect of incomplete knowledge of model parameters on decisional-relevant outcomes despite the wealth of available methods in the engineering literature.\cite{lee2009comparative}
First, the best practice prescribes a set of default probability distributions that are mainly driven by the consideration of the parameter's support.
For example, a beta distribution is used for characterizing the uncertainty of a parameter with a support $[0,1]$.
As a result, modelers tend to rely on "off-the-shelf" probability distributions to portray uncertainty "realistically over the theoretical range of the parameter."\cite{briggs2012model}
The use of "default distributions" is, in fact, a matter of convenience because there is no sure way to verify that our choice of the distribution and its parameters is indeed valid.
Furthermore, forcing the modelers to commit to a particular distribution implicitly assumes that the modelers have more information (e.g., knowing the shape of a distribution) than they actually possess and the uncertainty is known and quantifiable by a probability distribution.
Secondly, the best practice proposes the use of expert knowledge elicitation \cite{ohagan2006uncertain} if no prior data is available.
However, the proposed approach also hinges on a rather unverifiable assumption: the precise form of the probability distribution.
The lack of methodological guidance is due to the lack of an available approach for representing and propagating parameter uncertainty in situations where it is impossible to assume a precise probability distribution.
\\ \\
An ideal approach to parameter uncertainty characterization is one that requires minimal assumptions.
Specifically, in the absence of individual patient data, such an approach should require only information on statistics that are typically accessible to practitioners, such as mean, median, quantiles, minimum, and maximum (hereinafter collectively termed as minimal data).
Additionally, the ideal method does not require information on or assumptions about the precise form of a probability distribution.
Probability bounds analysis (PBA),\cite{ferson2015constructing} a combination of interval analysis and probability theory, is one such method and has been applied in risk engineering and management studies \cite{liu2017efficient, beer2013imprecise} and many other fields.\cite{enszer2011probability,kriegler2005utilizing,nong2007estimation}
In a PBA, the uncertainty about the probability distribution for each model parameter is expressed in terms of upper and lower bounds on the cumulative distribution function (CDF). 
These CDF bounds form a \textit{probability box} and are sufficient for circumscribing the unknown CDF of the model parameter given some minimal data about it.
The goal of this paper is to introduce the PBA method for representing and propagating parameter uncertainty in situations where knowledge or data about the parameter is limited and a probability distribution can not be specified precisely or the practitioners are not willing to commit to a particular form.
In this study, we assume that the model parameters are mutually independent.
This paper is organized into five main parts.
First, we review the concept of parameter uncertainty quantification.
Second, we formally describe an approach for representing parameter uncertainty in PBA using a probability box.
We focus on \textit{free probability boxes} that is a generalization of parametric probability boxes.
Then, we describe an approach for propagating probability boxes into a mathematical model.
Fourthly, we introduce an approach for decision-making using PBA results.
Then, we demonstrate two applications of PBA in modeling using Markov cohort models and a cost-effectiveness analysis of novel health technology.
Lastly, we conclude with a discussion on the limitations and directions for future research.
Throughout this exposition, we try to strike a balance between mathematical rigor and accessibility to practitioners.
\section{Preliminaries}\label{sec:prem}
We begin by briefly introducing the concept of parameter uncertainty quantification and the status quo approach of probability sensitivity analyses.
\subsection{Parameter uncertainty quantification} \label{sec:PUQ}
Let $\mathcal{M}$ denote a mathematical model (e.g., a cost-effectiveness model \cite{sanders2016recommendations} or a decision-analytic model \cite{iskandar2018theoretical}) that maps a set of $k$ inputs $\theta_i \in \bu{\theta}$ ($i=1,2,\ldots,k$) to a set of quantities of interest $\bu{Y}$, i.e.,  $\mathcal{M}: \bu{\theta} \rightarrow \bu{Y}$.
We treat $\mathcal{M}$ as a black-box model, i.e., only $\bu{\theta}$ and the corresponding $\bu{Y}$, after "running" $\mathcal{M}$ at particular values of or a realization of $\bu{\theta}$, are accessible. 
We assume that the values of $\bu{\theta}$ cannot be determined precisely due to lack of knowledge or data (\textit{epistemic uncertainty}). 
Our uncertainty about each parameter in $\bu{\theta}$ may vary according to the extent of what is known.
To quantify the effect of not knowing the values of parameters precisely on decision-relevant outcomes $\bu{Y}$ (parameter uncertainty quantification), we proceed with the following tasks. 
First, we specify a mathematical framework to encode the degree of uncertainty in the model parameters (parameter uncertainty representation). 
Then, we prescribe an approach to propagate parameter uncertainty, given a representation from the previous step, into our health economic model (parameter uncertainty propagation). 
Lastly, we set an approach to interpret the resulting uncertainty in the model outcomes for use in further analyses.
\subsection{Probabilistic sensitivity analyses} \label{sec:PSA}
If we adopt the standard approach for parameter uncertainty quantification, i.e.,  probabilistic sensitivity analysis (PSA),\cite{doubilet1985probabilistic} we proceed with the following steps.
For parameter uncertainty representation, we treat each parameter in $\bu{\theta}$ as a random variable that is endowed with a cumulative distribution function (CDF), $F(\theta)$, which is a monotonically increasing function from a sample space (e.g., the real number line $\mathbb{R}$) onto $[0,1]$, zero at negative infinity, and one at positive infinity.
In situations where the availability of data informing the estimation of the parameters is limited or non-existent, practitioners tend to select a type of CDF whose support matches with the model parameter's support (e.g., gamma distribution for non-negative parameters).
Hence, this common practice implicitly assumes that the form of $F(\theta)$ \textit{can be precisely specified}.
The location and ancillary parameters of the chosen distributions are typically estimated using a moment matching approach.
After the CDF has been assigned to each uncertain parameter, the uncertainty propagation follows an iterative Monte Carlo sampling approach. 
For each iteration, parameter values are sampled independently from the precisely specified CDFs, and the model is evaluated using these values to generate model outcomes. 
After a prespecified number of samples, an empirical CDF of the model outcome is obtained. 
Given the empirical distribution of an outcome, we can calculate its expected value and use it as an input to other analytical tasks (e.g., decision rule and value of information analysis).
\section{Parameter uncertainty representation}\label{sec:PUR}
This section introduces the parameter uncertainty representation step of PBA.
First, we describe the concept of a probability box.
Then, we introduce the formulas for a probability box given varying levels of available minimal data.
\subsection{Probability box}
As above, we suppose that the imperfect or lack of knowledge about a parameter ($\theta$) can be characterized by a random variable endowed with a CDF $F(\theta)$.
Instead of being restrictive in the context of limited data, PBA assumes that $F(\theta)$ is unknown or cannot be precisely specified and introduces the concept of a probability box or \textit{p-box}.
A probability box  of a continuous random variable $\theta$ with an unknown CDF $F(\theta)$ is an interval, $\mathcal{P}_{\mathcal{D}}=\left[\lF,\uF\right]$, which consists of all CDFs, including the unknown $F(\theta)$, that are: 1) bounded by a pair of bounding functions, i.e., a lower-bounding function (LBF) $\lF$ and an upper-bounding function (UBF) $\uF$ and 2) consistent with a minimal data $\mathcal{D}$ (where $\mathcal{D}$ denotes the set of available data or information on the statistics of the unknown CDF).\cite{ferson2015constructing,williamson1990probabilistic}
The UBF and LBF have the following properties:
\begin{enumerate}
    \item $\lF$ and $\uF$ are CDFs
    \item $\lF \le F(\theta) \le \uF$ for $\forall \theta$ in the support of $F(\theta)$
    \item $\lF$ and $\uF$ form the "tightest" bounds
    \item $\lF$ and $\uF$ are consistent with $\mc{D}$
\end{enumerate}
We say that a CDF of $\theta$ is consistent with the minimal data $\mc{D}$ if each element in $\mc{D}$ can be equated to a statistic that is derivable from the CDF.
Under the PBA framework, the epistemic uncertainty is given by: for every possible realization of $\theta$, we can only assign an interval of CDF values,  $\left[\lF,\uF\right]$, in contrast to a single CDF value.
As we accumulate more and more data on the parameter, the epistemic uncertainty is reduced, and the interval will eventually shrink to a single CDF.
\subsection{P-box formulas for different $\mathcal{D}$} \label{sec:pbox_formulas}
\begin{figure}[bt]
\centering
\includegraphics[width=14.5cm]{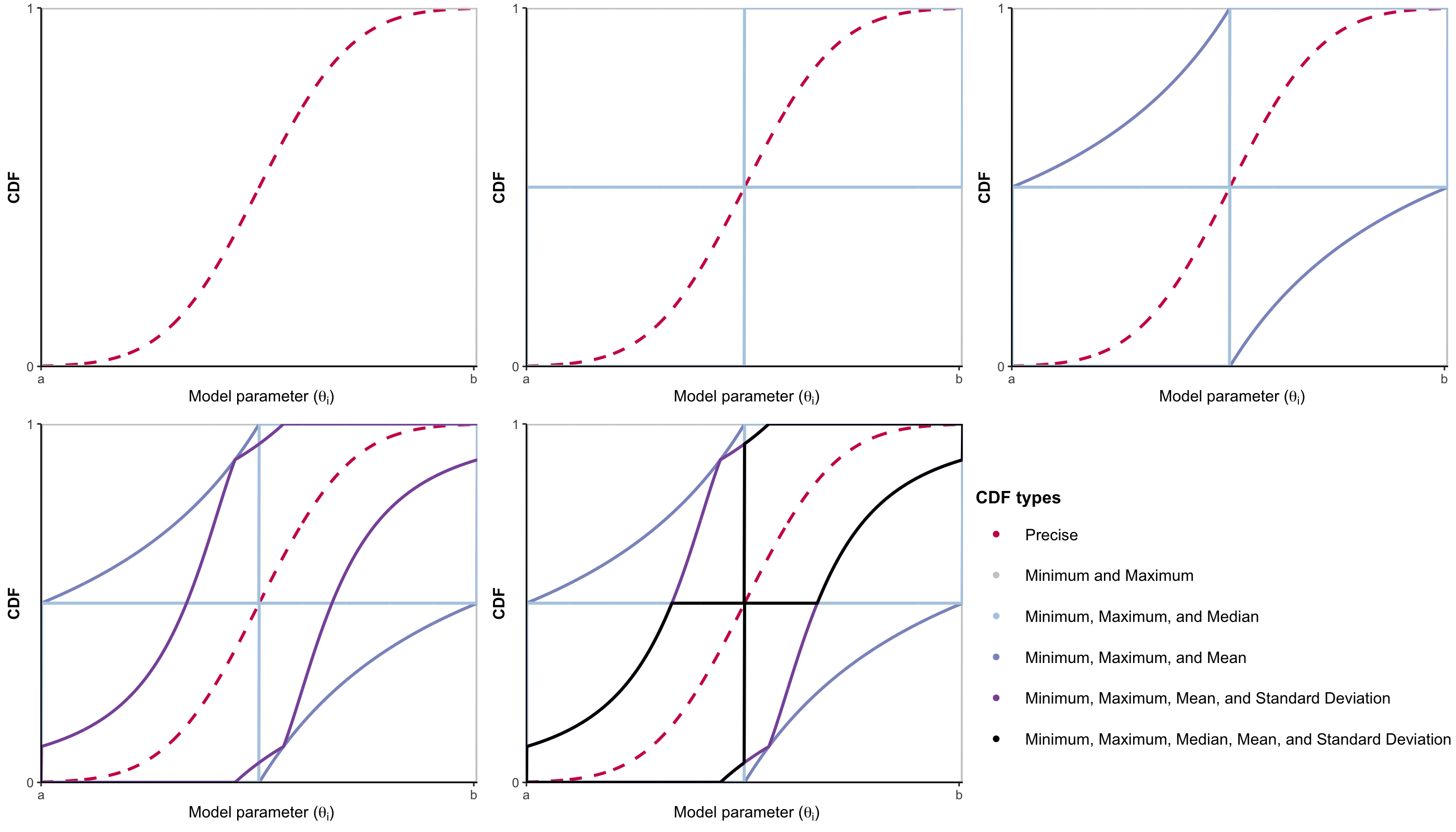}
\caption{P-boxes (solid) with different $\mathcal{D}$s and a normal CDF (dashed). CDF: cumulative distribution function.}
\label{fig:pbox_input}
\end{figure}

We consider commons situations of data availability where a modeler can identify and specify a combination of different summary statistics of and/or information on $\theta$ that constitutes a particular minimal data $\mc{D}$.
We show the derivation of one formula (Equation \ref{eq:pbox_std_LBF}) as an exemplar in Appendix \ref{app:pbox_derivation}.
We also show one proof of a p-box providing the tightest bounds on the unknown CDF, among all other pairs of bounding functions, given $\mc{D}$ (Appendix \ref{app:pbox_tight}).
\\ \\
The first situation involves knowing the smallest (minimum) and largest (maximum) values of a parameter.
For some parameters, one can infer the range from the theoretical limits, such as zero to one for probability or utility parameters. 
In some cases, a modeler may ask domain experts to specify a range from their knowledge about the quantity in question.
In both cases, the task will set a minimum $\ltt$ and a maximum $\utt$ such that the value of a parameter lies in the interval $[\ltt,\utt]$. 
The p-box $\mathcal{P}_{\ltt,\utt}$ is given by:
\begin{align}
    \lF_{\ltt,\utt} = \begin{cases}
     0 \quad &\mbox{ for } \quad \theta < \utt \\
    1 \quad &\mbox{ for } \quad \utt \le \theta\\
    \end{cases}
\end{align}
for LBF, and, 
\begin{align}
    \uF_{\ltt,\utt} = \begin{cases}
     0 \quad &\mbox{ for } \quad \theta < \ltt \\
    1 \quad &\mbox{ for } \quad \ltt \le \theta\\
    \end{cases}
\end{align}
for UBF.
\\ \\
If, in addition to knowing $\ltt,\utt$, the median $m$ of $\theta$ is also known, then the p-box will be tighter than $\mathcal{P}_{\ltt,\utt}$.
The p-box $\mathcal{P}_{\ltt,\utt,m}$ is given by:
\begin{align}\label{eq: pbox_median_LBF}
    \lF_{\ltt,\utt,m} = \begin{cases}
     0 \quad &\mbox{ for } \quad \theta < m \\
    0.5 \quad &\mbox{ for } \quad m \le \theta < \utt\\
        1 \quad &\mbox{ for } \quad \utt \le \theta \\
    \end{cases}
\end{align}
for LBF, and, 
\begin{align}\label{eq: pbox_median_UBF}
    \uF_{\ltt,\utt,m} = \begin{cases}
     0 \quad &\mbox{ for } \quad \theta < \ltt \\
    0.5 \quad &\mbox{ for } \quad \ltt \le \theta < m\\
        1 \quad &\mbox{ for } \quad m \le \theta\\
    \end{cases}
\end{align}
for UBF. \\ \\
If, in addition to knowing $\ltt,\utt$, the mean $\mu=E[\theta]$ is also known, then the p-box $\mathcal{P}_{\ltt,\utt,\mu}$ is given by:
\begin{align} \label{eq:pbox_mean_LBF}
    \lF_{\ltt,\utt,\mu} = \begin{cases}
     0 \quad &\mbox{ for } \quad \theta < \mu \\
    \frac{\theta-\mu}{\theta-\ltt} \quad &\mbox{ for } \quad \mu \le \theta < b\\
    1 \quad &\mbox{ for } \quad b \le \theta\\
    \end{cases}
\end{align}
for LBF, and, 
\begin{align}\label{eq:pbox_mean_UBF}
    \uF_{\ltt,\utt,\mu} = \begin{cases}
    0 \quad &\mbox{ for } \quad \theta < a \\
     \frac{\utt-\mu}{\utt-\theta} \quad &\mbox{ for } \quad a \le \theta < \mu \\
    1 \quad &\mbox{ for } \quad \mu \le \theta\\
    \end{cases}
\end{align}
for UBF. \\ \\
If, in addition to $\ltt,\utt$ and $\mu$ , we have data on the standard deviation $\sigma$, then the p-box $\mathcal{P}_{\ltt,\utt,\mu,\sigma}$ is given by:
\begin{align} \label{eq:pbox_std_LBF}
    \lF = \begin{cases}
     0 \quad &\mbox{ for } \quad \theta < \xi_1 \\
    \frac{\sigma^2+(b-\mu)(\theta-\mu)}{(\utt-\ltt)(\theta-\ltt)} \quad &\mbox{ for } \quad \xi_1 \le \theta <\xi_2\\
        \frac{(\theta-\mu)^2}{(\theta-\mu)^2 + \sigma^2} \quad &\mbox{ for } \quad \xi_2 \le \theta < \utt\\
    1 \quad &\mbox{ for } \quad b \le \theta\\
    \end{cases}
\end{align}
for LBF, and, 
\begin{align} \label{eq:pbox_std_UBF}
    \uF = \begin{cases}
         0 \quad &\mbox{ for } \quad \theta < a \\
      \frac{\sigma^2}{(\mu-\theta)^2+\sigma^2}  \quad &\mbox{ for } \quad \ltt \le \theta < \xi_1 \\
    \frac{(\utt-\mu)(\utt-\ltt+\mu-\theta) - \sigma^2}{(\utt-\ltt)(\utt-\theta)} \quad &\mbox{ for } \quad \xi_1 \le \theta <\xi_2\\
       1 \quad &\mbox{ for } \quad \xi_2 \le \theta\\
    \end{cases}
\end{align}
for UBF, where $\xi_1 = \mu - \frac{\sigma^2}{b-\mu}$ and $\xi_2 = \mu + \frac{\sigma^2}{\mu-a}$.
\\ \\
In principle, as we have additional summary statistics on $\theta$ or more information about the unknown $F(\theta)$, the \ p-box becomes tighter (Figure \ref{fig:pbox_input}).
Explicit formulas for other $\mc{D}$s have a complex form (see Appendix \ref{app:pbox_median_mean} for an example where $\mc{D}=\{a,b,m,\mu\}$) and are generally difficult to derive.\cite{ferson1996whereof}
In general, we can derive further cases by intersecting the p-boxes of different $\mathcal{D}$s described above (termed as primitive p-boxes) by "tracing" the minimum (or maximum) of the intersection of the corresponding UBFs (or LBFs).
More formally, for each $\mathcal{D}_{d}$ (where $d$ indexes each combination of available data), the LBF and UBF are given by:
\begin{align}
    \lF=\max \lFFD{1} \lFFD{2} \ldots \lFFD{n}
\end{align}
and
\begin{align}
    \uF=\min \uFFD{1} \uFFD{2} \ldots \uFFD{n},
\end{align}
respectively.
\section{Parameter uncertainty propagation}\label{sec:PUP}
In a modeling study, we typically have heterogeneity in the amount of and indirectness or imprecision in the available data used to estimate $\bu{\theta}$.
In principle, each $\theta_i \in \bu{\theta}$, based on the data availability and the chosen representation of its uncertainty, falls into of the following subsets of $\bu{\theta}$: (1) $\bu{\theta}_f$ for parameters with fixed values (no uncertainty), (2) $\bu{\theta}_c$ for parameters known up to their precise CDFs, and (3) $\bu{\theta}_b$ for parameters known up to their $\mc{D}$s.
In some cases, practitioners may have access to information that is sufficient for specifying probability distributions of parameters ($\bu{\theta}_c$).
This section presents the parameter uncertainty propagation step of PBA in the context of $\bu{\theta}_b$ only and a mix of $\bu{\theta}_c$ and $\bu{\theta}_b$.
First, we describe the intuition behind the propagation and proceed with an algorithm.
\subsection{Propagating p-boxes}
\begin{figure}[bt]
\centering
\includegraphics[width=12cm]{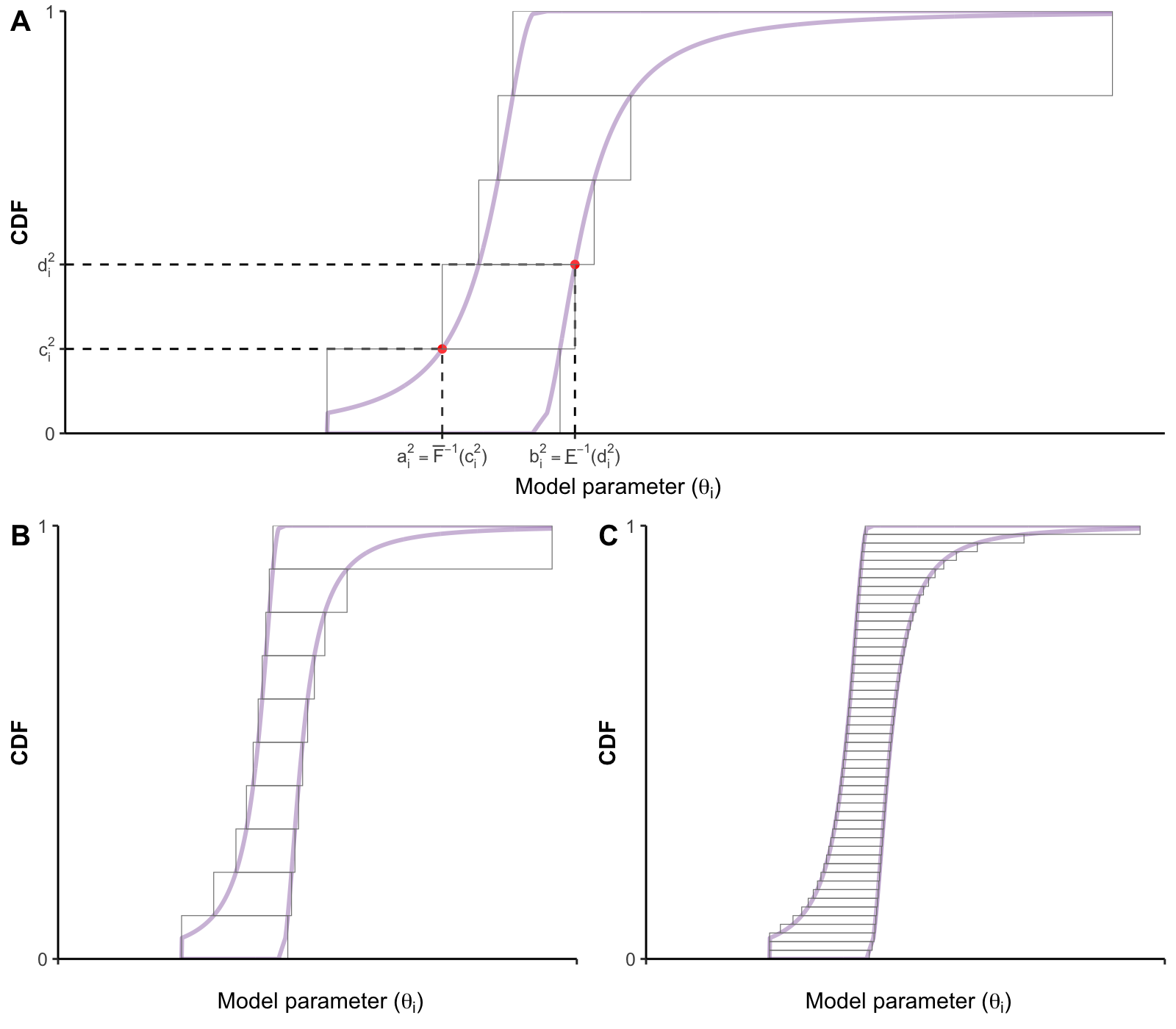}
\caption{Outer discretization approach for approximating the p-box of the model parameter $\theta_i$. Sub-figure A shows the sampling of an interval using the quasi-inverse of the p-box , given a particular sub-interval in $[0,1]$. Sub-figures B and C show the accuracy of the approximation when using, $n_{\theta_i}=10$ and $n_{\theta_i}=50$, respectively.}
\label{fig:pbox_approx}
\end{figure}

We recall that uncertainty propagation in PSA works as follows: each parameter value is sampled from its precise CDF, typically using the inverse transform sampling method if the inverse of the CDF is explicitly known.
For PBA, we use the same idea with one modification, i.e.,  we sample an interval of values instead of a single value.
The sampling scheme loosely mimics the inverse transform sampling.
For each $p \in [0,1]$ (the image of the CDF), we sample an interval by using the inverses of the LBF and UBF. 
To mitigate the computational burden, instead of sampling the intervals for all values in $[0,1]$, we partition the image of the CDF into finite sub-intervals. 
For each sub-interval and its endpoints, we calculate the corresponding interval of parameter values using the inverse p-box.
The choice of how to evaluate the endpoints of the sub-interval, e.g., using the LBF (UBF) for the upper (lower) endpoint, determines the accuracy of the approximation due to discretization.
We assign a probability to the interval based on the length of the sub-interval.
We then repeat the process for each parameter.
Since there are multiple possible realizations (equal to the number of sub-intervals) for each parameter, we need to consider all possible combinations of sub-intervals across all parameters, e.g., using a Cartesian product.
The probability of each possible combination (henceforth termed as a hyperrectangle) is computed by multiplying the probabilities assigned to the sub-intervals comprising the combination because of our independence among parameters assumption.
After specifying an approach for sampling hyperrectangles from p-boxes, we need to prescribe a method to evaluate our model using the sampled intervals.
One approach is based on optimization, where the goal is to find a pair of optima, i.e., the minimum and maximum values of the model outcome for each possible hyperrectangle.
The probability of observing a pair of optimum values is equal to the sampling probability of the corresponding hyperrectangle.
To obtain the p-box of the model outcome, we cumulate the probabilities of each minimum (maximum) to derive the UBF (LBF). 
\\ \\
Formally, to propagate the uncertainty of the set $\bu{\theta}_b$ into $\mc{M}$, we proceed with the following steps.
First, for each $\theta_i \in \bu{\theta}_b$, we derive its p-box given $\mc{D}_i$.
Next, we specify the approach for generating interval-valued samples.
To build intuitions about how the sampling works, we use a full-factorial-design approach, i.e., the slicing algorithm with outer discretization \cite{zhang2010interval}.
For each $\theta_i$, the interval $[0,1]$ is partitioned into $n_{\theta_i}$ sub-intervals with its corresponding probability mass $m_j$ ($j \in \{1, 2, \ldots,n_{\theta_i}\}$) such that $\sum\limits_{j=1}^{n_{\theta_i}} m_j = 1$.
For the $j$-th sub-interval, we identify its lower and upper boundary points, i.e., $c_i^j$ and $d_i^j$, respectively.
Given each pair of boundary points, we calculate the boundary points in the $\theta_i$ domain by using the inverse or quasi-inverse of the p-box:
\begin{align} \label{eq:outer}
    \ltt_i^{j} = \uFp^{-1}\left(c_i^j\right), \quad \quad \utt_i^{j} = \lFp^{-1}\left(d_i^j\right)
\end{align}
These inverse or quasi-inverse functions are derived from their corresponding LBF and UBF (Appendix \ref{app:pbox_inverses}), where the quasi-inverses are due to the fact that some LBFs and UBFs are not strictly injective functions.
Equation \ref{eq:outer} corresponds to a particular choice of a discretization, i.e., an outer discretization approach (Figure \ref{fig:pbox_approx}).
The intervals $[\ltt_i^{j},\utt_i^{j}]$ and their associated $m_j$ collectively form the p-box of $\theta_i$.
\\ \\
We denote $\mathcal{K}$ as the set of multi-indices where each multi-index corresponds to a particular combination of sub-intervals of all $\theta$s in $\bu{\theta}_b$:
\begin{align}\label{eq:K}
    \mathcal{K}=\left\{\bu{k} \vert \bu{k}=(k_1, k_2, \ldots, k_n), k_i \in \{1, \ldots, n_{\theta_i}\}\right\}
\end{align}
where $n$ is the number of parameters in $\bu{\theta}_b$ and $k_i$ indexes the sub-intervals for $\theta_i$.
For each $\bu{k} \in \mathcal{K} $, we denote $\mathcal{H}_\bu{k}$ as a hyperrectangle (i.e., a Cartesian product of intervals) which is given by:
\begin{align}\label{eq:H_k}
    \mathcal{H}_{\bu{k}} = \left[\ltt_1^{k_1},\utt_1^{k_1}\right] \times \left[\ltt_2^{k_2},\utt_2^{k_2} \right] \times \ldots \times \left[\ltt_n^{k_n},\utt_n^{k_n} \right]
\end{align}
For each $\mathcal{H}_{\bu{k}}$ we calculate its probability mass as follows:
\begin{align} \label{eq:H_k_probs}
    P(\mathcal{H}_{\bu{k}}) = m_{k_1}  m_{k_2} \ldots  m_{k_n}
\end{align}
where $m_{k_i}$ corresponds to the probability mass associated with the $k_i$-th subinterval of $\theta_i$.
Equation \ref{eq:H_k_probs} represents our assumption about the independence among model parameters since the dependence structure depends on how the probabilities in the Cartesian product are computed.
In our case, we multiply the marginal probabilities to get the probabilities of each $\mc{H}_\bu{k}$, which is akin to assuming random set independence.\cite{ferson2015constructing}
\\ \\
For each $\mathcal{H}_{\bu{k}}$, we associate two optimization problems whose solutions provide the bounds on a quantity of interest (model outcome) $Y$:
\begin{align} \label{eq:optim}
    \lY^{\bu{k}} = \min_{\bu{\theta}_b} \mathcal{M}(\bu{\theta}_b,\bu{\theta}_f,\bu{\theta}_c), \quad \quad \uY^{\bu{k}} = \max_{\bu{\theta}_b} \mathcal{M}(\bu{\theta}_b,\bu{\theta}_f,\bu{\theta}_c)
\end{align}
for a particular set of values of $\theta$s in $\bu{\theta}_f$ and  $\bu{\theta}_c$.
The existence of a maximum and a minimum is guaranteed by the Weierstrass Extreme Value Theorem \cite{rudin1964principles} since, in decision-analytic modeling, the model $\mathcal{M}$ is typically smooth and $\mathcal{H}_{\bu{k}}$ is closed and bounded (compact).
The p-box of $Y$ is therefore characterized by a collection of $\left[\lY^{\bu{k}},\uY^{\bu{k}}\right]$ and its corresponding probability mass $P(\mathcal{H}_{\bu{k}})$.
The empirical p-box of $Y$ can be calculated as:
\begin{align}\label{eq:cdf_Y}
 \underbar{$F$}(Y)=\sum\limits_{i=1}^{|\mathcal{K}|}  P(\mathcal{H}_{\bu{k}_i}) \mathcal{I}_{\lY^{\bu{k}_i} \le Y}   , \quad \quad \overline{Y}=\sum\limits_{i=1}^{|\mathcal{K}|}  P(\mathcal{H}_{\bu{k}_i}) \mathcal{I}_{\uY^{\bu{k}_i} \le Y} 
\end{align}
where $\bu{k}_i$ indexes all elements in $\mc{K}$, $|\mathcal{K}|$ denotes the number of elements in $\mc{K}$, and $\mathcal{I}_{\uY^{\bu{k}_i} \le Y}$ is an indicator function.
\subsection{Propagating p-boxes and precise CDFs} \label{sec:propagate_theta_c}
To propagate uncertainty from both sets $\bu{\theta}_c$ and $\bu{\theta}_b$ into $\mathcal{M}$, we proceed in two steps.
First, since the uncertainty of each parameter in $\bu{\theta}_c$ can be characterized by a precise CDF, the uncertainty propagation reduces to a Monte Carlo approach,\cite{cullen1999probabilistic} a repeated sampling from a joint distribution of parameters in $\bu{\theta}_c$ (if their dependencies are known).
Let $f(\bu{\theta}_c)$ be the joint distribution.
Repeated samplings from $f(\bu{\theta}_c)$ will generate a sequence of samples of $\bu{\theta}_c$: $\bu{\theta}_c^{1}, \bu{\theta}_c^{2}, \ldots, \bu{\theta}_c^{N}$, where $N$ is the total number of Monte Carlo samples. 
Second, for each sample  $\bu{\theta}_c^{l}$ ($l$ indexes the parameter in $\bu{\theta}_c$) and each $\mathcal{H}_k$ (Equation \ref{eq:H_k}), we solve the following optimization problems:
\begin{align} 
    \lY^{\bu{k},l} = \min_{\bu{\theta}_b} \mathcal{M}(\bu{\theta}_b,\bu{\theta}_f,\bu{\theta}_c^l), \quad \quad \uY^{\bu{k},l} = \max_{\bu{\theta}_b} \mathcal{M}(\bu{\theta}_b,\bu{\theta}_f,\bu{\theta}_c^l)
\end{align}
and derive $\underbar{$F$}^l(Y)$ and $\overline{F}^l(Y)$ using Equation \ref{eq:cdf_Y}.
The p-box of $Y$ is then calculated by averaging over the $N$ samples:
\begin{align}\label{eq:cdf_Y_theta_c}
    \widehat{\underbar{$F$}}(Y)=\sum\limits_{l=1}^{N}=\underbar{$F$}^l(Y), \qquad \widehat{\overline{F}}(Y)=\sum\limits_{l=1}^{N}=\overline{F}^l(Y)
\end{align}
Alternatively, if $\mc{M}$ is relatively linear, we can fix the values of $\theta$ in $\bu{\theta}_c$ at their mean values.
This approach avoids the use of repeated sampling and reduces the computational time.
\section{Application of PBA}
This section describes how practitioners can utilize the results of uncertainty propagation using PBA (Equations \ref{eq:cdf_Y} and \ref{eq:cdf_Y_theta_c}).
First, we introduce notations to fix ideas.
Then, we describe an application in decision analysis.
\subsection{Formalism of a decisional problem}
A typical decision-making problem in health domains consists of: 1) $m$ competing interventions (e.g., new drug vs. usual care), $a_r$ ($r=1,\ldots,m$); 2) $n$ decision-relevant outcomes (e.g., life expectancy or lifetime cost), $Y_j$ ($j=1,\ldots,n)$; 3) a mathematical model to evaluate the effect of $a$ on $\bu{Y}$ as in Section \ref{sec:PUQ}, $\bu{Y} = \mathcal{M}(a_r \vert \bu{\theta})$; 4) $k$ model parameters, $\theta_i$ ($i=1,\ldots,k$); 5) measures of knowledge or uncertainty about each parameter (e.g., precise CDFs or p-boxes) and their dependencies; 6) a value (or utility) function, $U(\mathcal{M} (a_r \vert \bu{\theta})):=U(a_r\vert \theta)$, that integrates the evaluation of each intervention on all $Y_j$s; and 7) a choice function capturing a decision rule for choosing the (or set of) optimal intervention(s), $G(U(\bu{a}\vert \theta))=\mathcal{A}$, where $\mc{A}$ is the set of optimal interventions.
For ease of exposition and without loss of generality, we assume that $\mc{M}$ is deterministic.
Hence, the states of the world are completely determined by our knowledge about $\bu{\theta}$.
\subsection{Decision analysis with PBA}
We recall that the most commonly used decision rule, i.e., Expected Value Maximization, requires the specification of CDFs in the context of parameter uncertainty.\cite{gilboa2009theory}
Under this choice function, if we can specify all the CDFs $F_i(\theta_i)$, then an intervention $a^*$ is chosen if
\begin{align}\label{eq:EU}
    a^* = \argmax_{a_r}\left\{\int_{\bu{\theta}} U(a_r\vert \bu{\theta})dF(\bu{\theta})\right\}
\end{align}
Since the propagation of uncertainty in $\bu{\theta}$ results in uncertainty in $\bu{Y}$ ($F(\bu{Y})$), we write $a^* = \argmax_{a_r}\left\{ \int_{\bu{Y}}  U(a_r \vert \mathbf{Y}) dF(\bu{Y})\right\}$.
We note that the calculation of the expected value of $\bu{Y}$ over its p-box results in an interval of expected values. 
The interval includes all expected values that correspond to CDFs enclosed by the p-box.
This is true because the p-box of $\bu{Y}$ is guaranteed to enclose all CDFs of $\bu{Y}$  (assuming that the p-boxes of the $\theta$s are properly specified).
Therefore, the expected value of $U(a_r \vert \bu{Y})$ for each CDF in the p-box must lie in the interval that is given by $[\underbar{$\mu$}(a_r),\overline{\mu}(a_r)]$ where $\underbar{$\mu$}(a_r)$ and $\overline{\mu}(a_r)$ are given by $\int_{\bu{Y}} U(a_r \vert \mathbf{Y}) d\underbar{$F$}(\bu{Y})$ and $\int_{\bu{Y}} U(a_r \vert \mathbf{Y}) d\overline{F}(\bu{Y})$, respectively.
Given what we know and assume about the uncertainty of the model parameters, the expected utilities can not be larger (smaller) than $\overline{\mu}(a_r)$( $\underbar{$\mu$}(a_r)$).
Furthermore, $[\underbar{$\mu$}(a_r),\overline{\mu}(a_r)]$ is not endowed with an uncertainty measure, i.e., we cannot say the relative plausibilities of each value in the interval.
Therefore, we cannot use the Expected Value Maximization for PBA.
Instead, the decision rule is based on finding the optimal intervention by comparing the intervals $[\underbar{$\mu$}(a_r),\overline{\mu}(a_r)]$ of all interventions ($a_r$).
\\ \\
Suppose that we have two competing interventions, i.e., $a_1$ and $a_2$, with their corresponding intervals $[\underbar{$\mu$}(a_1),\overline{\mu}(a_1)]$ and $[\underbar{$\mu$}(a_2),\overline{\mu}(a_2)]$ , respectively.
We conclude that $a_1$ is preferred to $a_2$ if:
\begin{description}
\item[Dominance] $\underbar{$\mu$}(a_1) > \underbar{$\mu$}(a_2)$ and $\overline{\mu}(a_1) > \overline{\mu}(a_2)$
\item[Pessimist] $\underbar{$\mu$}(a_1) > \underbar{$\mu$}(a_2)$
\item[Optimist] $\overline{\mu}(a_1) > \overline{\mu}(a_2)$
\item[Hurwicz criterion] $ \alpha\underbar{$\mu$}(a_1) + (1-\alpha) \overline{\mu}(a_1) >  \alpha\underbar{$\mu$}(a_2) + (1-\alpha) \overline{\mu}(a_2)$ 
\end{description}
The $\alpha$ in the Hurwicz decision criterion captures a decision-maker's relative attitude towards being overly pessimistic.
The choice of the decision rule is decisional-problem dependent and is typically driven by the the type of outcomes and the decisionmaker's risk preference.
\section{Case studies}
We conduct two case studies. 
The first case study uses a hypothetical Markov cohort model to examine the characteristics of PBA and demonstrate the difference between PBA and PSA.
The second case study is based on a published early assessment of the cost-effectiveness of a computer-assisted total knee replacement in the absence of clinical trial data.\cite{dong2006early} 
The models are coded in R\cite{R2019} and available under a GNU GPL license and can be found at https://github.com/rowaniskandar/PBA.
\subsection{Case study 1}
\begin{figure}[bt]
\centering
\includegraphics[width=9cm]{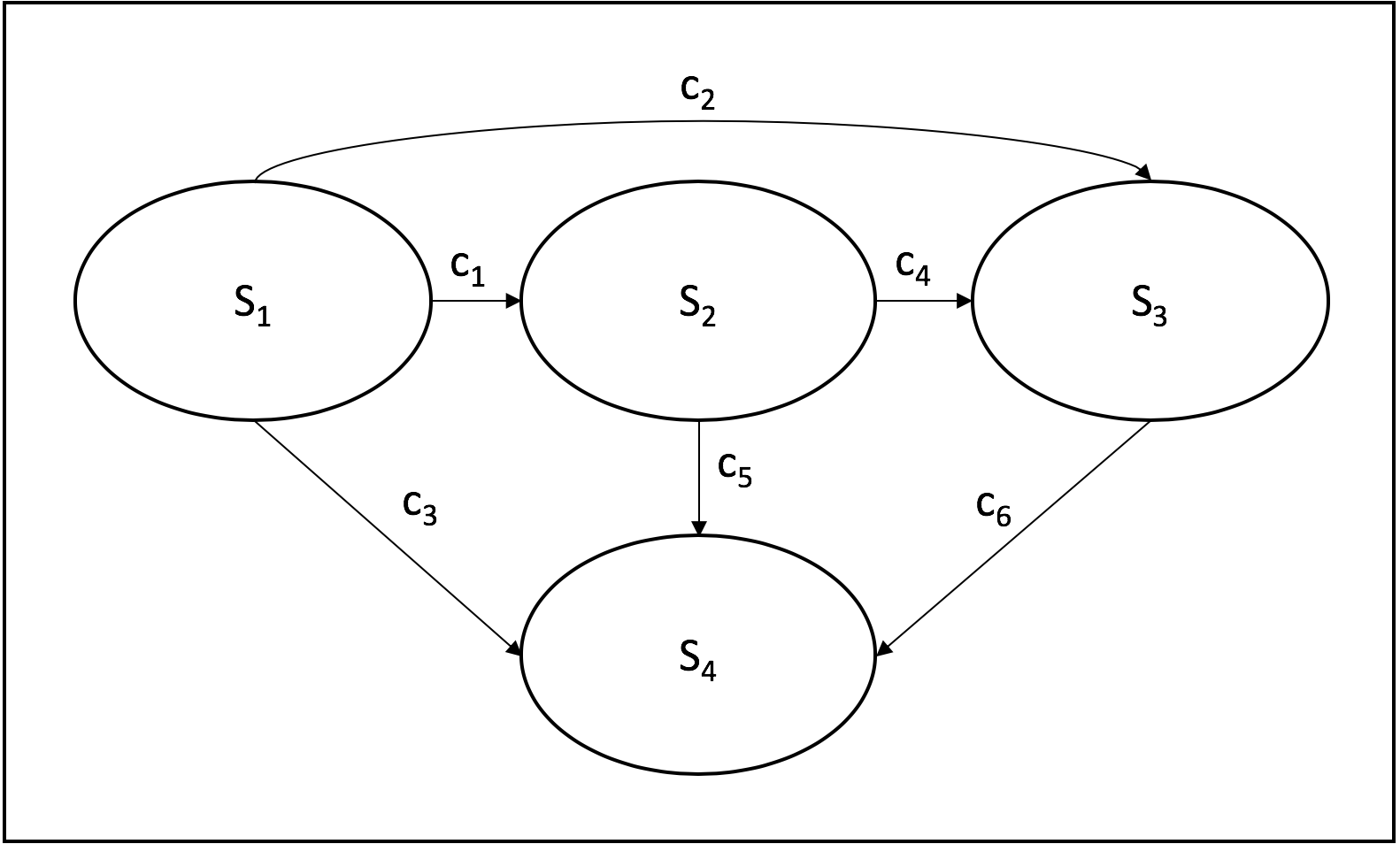}
\caption{A state-transition model diagram used in case study 1}
\label{fig:4statesdiagram_2}
\end{figure}

\begin{figure}[bt]
\centering
\includegraphics[width=14cm]{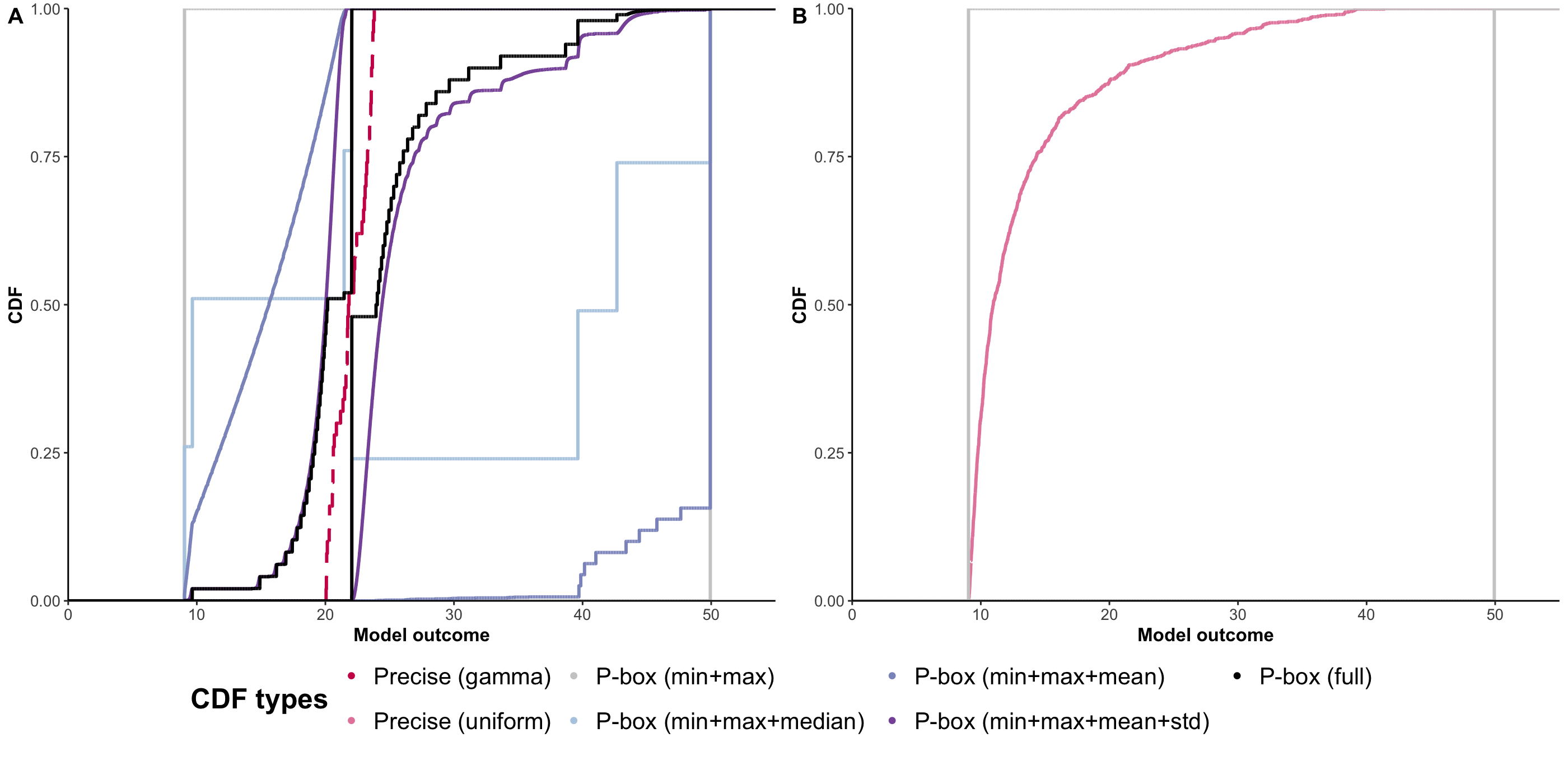}
\caption{Uncertainty around model outcome of $4$-state model using p-boxes vs. precise CDFs for $c_1$ and $c_6$. 
Sub-figure A portrays the comparison of the uncertainties in the model outcome resulting from a p-box and a gamma distribution.
Each (.) corresponds to the different combination of available  data on parameter ($\theta_b$).
As more information is available, the p-box enclosing the unknown precise CDF becomes tighter.
Sub-figure B illustrates the comparison between a p-box and a uniform distribution and demonstrates how p-box is more honest in representing the uncertainty, given information only on the minimum and the maximum values of the model parameters.
 CDF: cumulative distribution function.}
\label{fig:pbox_QoI}
\end{figure}

We consider a generic four-state stochastic (Markov) cohort model as our $\mc{M}$, which is commonly used in DAM and CEA studies,\cite{hunink2014decision} with the following health states $\mathbb{S}=\{S_1, S_2, S_3, S_4\}$ (Figure \ref{fig:4statesdiagram_2}), where $S_4$ is an absorbing state.
We assume that the probability distributions for the rates of transitions $S_1 \rightarrow S_3$ ($c_2$), $S_2 \rightarrow S_3$ ($c_4$), and $S_2 \rightarrow S_4$ ($c_5$) are known or that our knowledge is sufficient for precise specifications of CDFs.
We fix the values of the four parameters at their mean values: $c_2=0.01$, $c_3=0.001$, $c_4=0.1$, and $c_5=0.05$.
For $c_1$ (rates of transitions $S_1 \rightarrow S_2$) and $c_6$ (rates for $S_3 \rightarrow S_4$), we conduct two sets of comparisons. 
First, we compare the following scenarios: 1) the PBA scenario where the uncertainties in $c_1$ and $c_6$ are modeled using p-boxes with $\mathcal{D}=\{a=0,b=10,\mu=0.05,\sigma=0.00033\}$ and $\mathcal{D}=\{a=0,b=10,\mu=1,\sigma=0.0167\}$, respectively, and 2) the uncertainties in both rates are precisely specified using a gamma distribution with the same $\mathcal{D}$s as in the PBA scenario.
For the latter scenario, the uncertainty propagation follows the PSA approach.
This comparison demonstrates the effect of different degrees of conservatism, i.e., precise vs. imprecise CDFs, on the resulting uncertainty in the model outcome.
For the second comparison, we assume that only the minimum and maximum values of $c_1$ and $c_6$ are available to illustrate how PBA treats extreme data sparsity with fewer assumptions when compared to the common practice of using uniform distributions.
For the model outcome of interest, we calculate the expected residence time in states other than $S_4$ (Figure \ref{fig:pbox_QoI}).
For uncertainty propagation with precise CDFs, we use the support point method\cite{mak2018support} for sampling from the gamma and uniform distributions with $N=50$.
For uncertainty propagation using p-boxes (Equation \ref{eq:cdf_Y}), we apply a deterministic search algorithm based on systematic divisions of the domain (Equation \ref{eq:H_k}) into smaller hyperrectangles\cite{gablonsky2001locally} and use the implementation of the \textit{nlopt} library\cite{johnson2014nlopt} in the R program.\cite{ypma2014introduction}
\begin{figure}[bt]
\centering
\includegraphics[width=14cm]{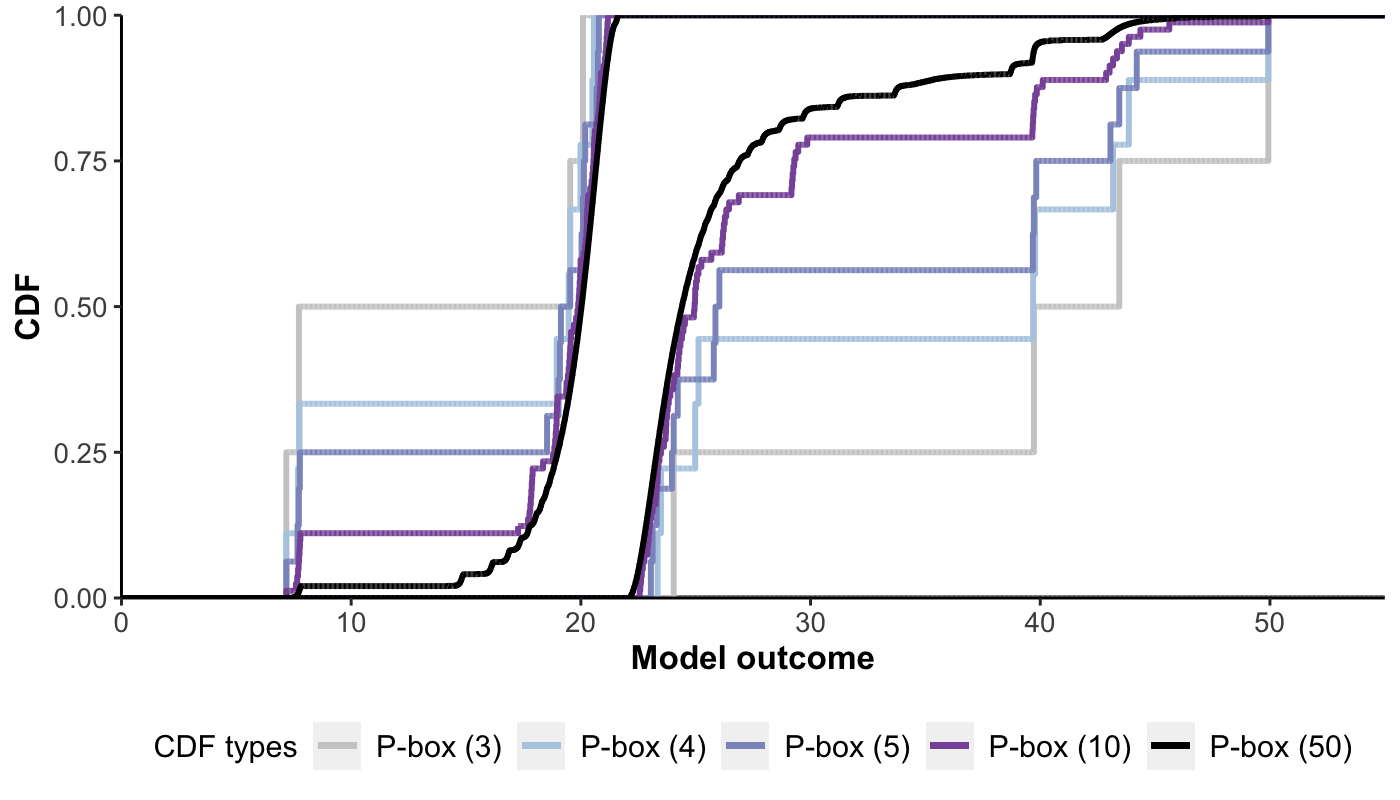}
\caption{The accuracies of the approximations of the p-box of the model outcome as a function of the increasing numbers of sub-intervals (as indicated by the numbers in the parentheses) for each parameter in $\theta_b$, given data on $a$, $b$ $\mu$, and $\sigma$. CDF: cumulative distribution function.}
\label{fig:pbox_comparison_multiple_pbox_interval}
\end{figure}
\\ \\
The first comparison shows the difference between the results of a parameter uncertainty propagation into a model outcome using precise CDFs (gamma distribution) vs. p-box.
A PBA results in a p-box enclosing the unknown CDF of the model outcome instead of a precise CDF (Figure \ref{fig:pbox_QoI}A).
The p-box gives additional information: (1) the amount of uncertainty in the model outcome due to our imperfect or complete lack of knowledge about some model parameters, which is indicated by the area enclosed by the p-box and (2) the plausible values of the model outcome, which is indicated by the model outcome values with non-zero probabilities.
The latter suggests the minimum and maximum achievable values of the model outcome.
We also note that the accuracy of the empirical LBF and UBF increases with the number of sub-intervals of each parameter ($n_{\theta_i}$).
The second comparison showcases the implications of how uncertainty due to a severe lack of data about parameter values is modeled (Figure \ref{fig:pbox_QoI}B).
Uncertainty propagation with a uniform distribution results in a model outcome's CDF that gravitates towards a central tendency and, essentially, "eliminates" our ignorance.
In contrast, the result of PBA preserves our ignorance.
Furthermore, we observe that the plausible values of the model outcome under uniform distributions are concentrated in the leftmost region of the support, thereby discounting the possibility of having high values.
Conversely, PBA produces bounds on the model outcome while maintaining the plausibility of a wide range of values.
This observation highlights the potential peril of assuming a precise form of a CDF, particularly when the model outcome represents an undesirable outcome.
\subsection{Case study 2}
\begin{figure}[bt]
\centering
\includegraphics[width=14.5cm]{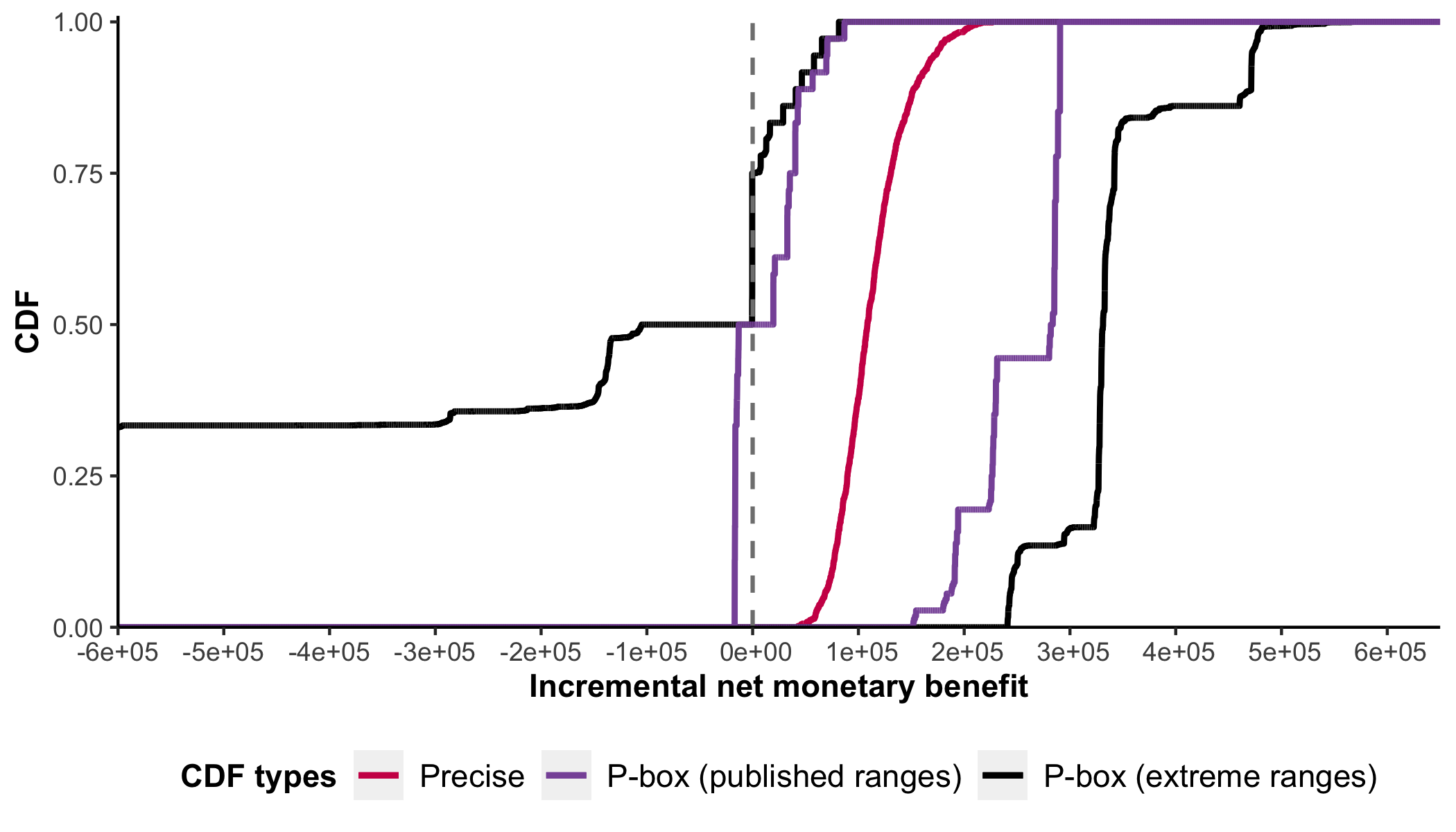}
\caption{Uncertainties around the incremental net monetary benefit of computer-assisted vs. standard total knee replacement surgeries using (1) precise CDFs, (2) p-boxes with published minimum and maximum values, and (3) p-boxes with extreme minimum and maximum values. The dashed vertical line lies at the zero INMB. CDF: cumulative distribution function.}
\label{fig:INMB}
\end{figure}

We replicate a published cost-effectiveness analysis of a computer-assisted total knee replacement (CA-TKR) versus a conventional TKR.\cite{dong2006early}
We develop a Markov model with the following states:  TKR operation for knee problem, normal health after primary TKR, TKR with minor complications, TKR with serious complications, simple revision operation for treating complications, complex revision operation for treating complications, other non-revision treatments for complications, normal health after TKR revision, and death.
The analytical period is ten years with a monthly time-step.
For the transition probabilities which could not be estimated from available data, i.e., transitions to serious complication from minor complication or other treatment, transitions to minor complication from other treatment or serious complication, and transitions to simple revision from other treatment or vice versa, the authors assumed that their values are identical to the estimated mean values for the same transitions from other states.
We relax these assumptions and, instead, subject the six transition probabilities and the efficacy of CA-TKR to an uncertainty analysis using PBA and PSA with data only on the mean, minimum, and maximum values.
For probabilities and the efficacy parameters, we use beta and gamma distributions, respectively.
Since the study does not report the variances, we assume that the standard deviation is $20\%$ of the mean value.
We conduct two uncertainty analyses in which we vary the minimum and maximum values (ranges) of the seven parameters of interest, i.e., the ranges reported in the study and wider ranges of values (ten times the original ranges).
For the other parameters, we fix them at their mean values (see Table 2 in \cite{dong2006early}).
For the cost-effectiveness measure, we calculate the incremental net monetary benefit (INMB) and estimate its empirical CDF (PSA) and p-box (PBA), given a willingness-to-pay threshold of \pounds $30000$ per quality-adjusted life year.
The cost-effectiveness analysis is conducted from the National
Health Services' perspective and uses $3.5\%$ as the discount rate.
We use all the data and assumptions that are reported in the study and make reasonable assumptions whenever the data is not available in the published paper.
For more details on the model structure and estimation and their assumptions, we refer the readers to the original study \cite{dong2006early}.
\\ \\
Using the PSA approach where we assume precise specifications of the CDF and use the published ranges, the CDF of the INMB lies entirely to the right of zero in Figure \ref{fig:INMB}, i.e.,  CA-TKR is always cost-effective at the given willingness-to-pay threshold.
In contrast, the PBA approach using the same ranges results in a p-box of the INMB, a marginal part of which is to the left of zero line, i.e., CA-TKR is not cost-effective at the given willingness-to-pay threshold.
If we consider a wider range of possible values for each of the seven parameters, the p-box stretches to, at least, \pounds $-600000$.
This observation indicates that the uncertainty in the INMB is sensitive to the assumed ranges of values.
Therefore, the cost-effectiveness of CA-TKR is overestimated when we assume rather narrow ranges for model parameters for which we lack reliable data.
\section{Discussion}
This study introduces the probability bound analysis method for quantifying the effect of parameter uncertainty on decision-relevant outcomes that is distribution-free.
This paper is the first study that examines the utility of PBA in DAM and CEA studies.
Although our contribution focuses mainly on medical decision-making and economic evaluation fields, the methodologies apply to many studies using mathematical models to inform policy decisions.\cite{den2019guidelines}
To assist practitioners, we provide p-box formulas for the most common situations of data availability.
We show an approach for propagating p-boxes into a black-box model where the uncertainty of the model parameters is characterized by a combination of p-boxes and precise distribution functions.
We conduct two case studies to demonstrate the methodological characteristics and practical application of PBA.
\subsection{Advantages of PBA}
The novel approach allows practitioners to conduct probabilistic assessments even when extremely little reliable empirical information is available about the distributions of model parameters.
In PBA, parameter uncertainties are characterized by p-boxes that provide the maximum area of uncertainty (tightest bounds) containing the unknown distribution function, given knowledge about the summary statistics of the parameters.
For basic binary operations (addition, subtraction, multiplication, and division),\cite{ferson2004arithmetic,williamson1990probabilistic} the derived p-box of a model outcome is optimal in the following sense: one can not find other tighter bounds without excluding some of the plausible CDFs.
However, p-box computations using basic operations cannot be easily extended to black-box models.\cite{aughenbaugh2007probability}
Nevertheless, the uncertainty propagation of p-boxes into a black-box model, using optimization (Equation \ref{eq:optim}), generates bounds that are guaranteed to enclose \textit{all possible CDFs} of the model outcome provided that the parameter p-boxes enclose their respective distributions, without the assurance of the optimality of the bounds \cite{beer2013imprecise} .
PBA is based on two existing approaches. 
First, we believe that a parameter value can be bounded in some intervals without specifying the relative plausibility over the interval (\textit{interval analysis} \cite{alefeld2012introduction}).
Secondly, we assert that the parameter uncertainty can be represented by a probability (\textit{probability theory} \cite{savage1972foundations}).
When taken together, PBA models the uncertainty using a CDF, but the CDF is not precisely specified and assumed to be located within an interval containing all possible CDFs.
In a way, a PBA gives an identical answer as an interval analysis whenever the range is the only accessible information. 
If the lower and upper bounds of a CDF coincide for every element in the support, then a p-box degenerates to a CDF: a situation where Monte Carlo simulation is the standard approach.
Therefore, a PBA is a generalization of the two standard approaches for representing parameter uncertainty.
In addition, a PBA is an improvement over both approaches for situations where one approach is not sufficient by itself.
\\ \\
One decision-relevant information from the results of a PBA, as demonstrated in our case studies, is the bounds on the plausible values of a model outcome.
This information is particularly useful when the model outcome represents a negative outcome (or a catastrophic event).\cite{tucker2003probability}
The p-box of a model outcome suggests that the outcome will not be smaller (or larger) than a minimum (or maximum), which can be identified by the infimum (supremum) of the support of the UBF (LBF). 
The standard approach in DAM and CEA studies, i.e., the (over-)reliance on using "off-the-shelf" probability distributions for characterizing uncertainties about model parameters, may potentially lead to an underestimation of the probability of observing extreme values of the model outcome. 
Using probability distributions may also assume more information about uncertainty than that is supported by the current evidence base.
These errors in estimating probabilities in the context of insufficient data or a complete lack of knowledge may contribute to overconfidence and lead to a failure to insure ourselves against highly consequential risks.\cite{taleb2007black}
Our first case study also highlights the consequence of using a uniform distribution, the most common approach for modeling ignorance about a parameter.
Although using a uniform distribution may be justifiable as the embodiment of the principle of indifference \cite{norton2008ignorance, laplace1812marquis}, this "all are equally likely" assumption significantly discounts the possibility of the extremes. 
On the other hand, PBA can, loosely speaking, transfer our ignorance about parameter values to ignorance about a model outcome.
The second case study \cite{dong2006early} represents a real-world setting where we lack the data to inform some of the key parameters, including the efficacy of the novel technology and probabilities of adverse events.
The authors of the original study failed to adequately represent the uncertainties in these parameters by prescribing narrow ranges.
In our re-analysis, our PBA approach yields a wider p-box of the INMB (more uncertainties) when assuming wider ranges of values.
Moreover, the PBA does not require any assumptions about the standard deviations (cf. Equations \ref{eq:pbox_mean_LBF} and \ref{eq:pbox_mean_UBF}).
Although we are not able to exactly replicate the published results due to missing information on the variances, our qualitative result is still valid.
Regardless of the value of the variances, the conclusion on whether CA-TKR is cost-effective is sensitive to the assumed ranges of values.
\subsection{Computational costs}
PBA is computationally intensive for the following reasons. 
First, an implementation of a PBA requires an optimization step over the p-box.
In this study, we use a full factorial design that transforms the problem of propagating a p-box into propagation of a large number of intervals.
The higher the required level of accuracy is, the higher number of intervals $n_{\theta_i}$ is needed.
Furthermore, an increase in the number of p-box parameters will lead to a higher-dimensional optimization problem.
Second, the computational burden is further exacerbated if the black-box model ($\mathcal{M}$) is "expensive" to evaluate for a given $\bu{\theta}$.
Thirdly, if, in addition to p-box parameters, some parameters are characterized by their CDFs, the optimization step is embedded in a Monte Carlo sampling loop (Section \ref{sec:propagate_theta_c}); thereby increasing the number of optimizations by a factor of $N$ (the total number of Monte Carlo samples).
To mitigate the computational burden, users of PBA may opt to use less conservative p-box propagation approaches \cite{schobi2017uncertainty}, more efficient optimization methods\cite{deng2013improved}, and fast-to-evaluate approximations of the original model or meta-models.\cite{ellis2020active}
Nevertheless, we expect a higher computational burden since a PBA imposes fewer restrictions (i.e., we do not assume a functional form), leading to a larger region of uncertainty over which a model needs to be evaluated.
\subsection{Relation to other methods}
PBA is generally regarded as one of the uncertainty quantification approaches related to the theory of imprecise probability.~\cite{augustin2014introduction,beer2013imprecise}
In particular, a p-box is closely connected to Dempster-Shafer’s theory of evidence.\cite{dempster2008upper,shafer1976mathematical,ferson2015constructing}
The LBF and UBF can be interpreted as belief and plausibility measures for the event $\theta$ taking values than a particular value $\{\theta \le x\}$. \cite{ferson2015constructing}
In Dempster-Shafer’s theory, the belief function describes the minimum amount of probability that must be associated with the event, whereas the plausibility function describes the maximum amount of probability that might be associated with the same event.
The PBA framework is also related to Bayesian sensitivity analysis (or robust Bayesian analysis).\cite{berger1990robust}
In this approach, an analyst’s uncertainty about which prior distribution and likelihood function should be used is characterized by an entire class of prior distributions and likelihood functions. 
The analysis proceeds by studying various outcomes for each possible combination of prior distribution and likelihood function. 
Another distribution-free approach is the Chebyshev inequality\cite{saw1984chebyshev} that can be used to compute bounds on the CDF of a random variable, given the mean and standard deviation of the random variable.
However, the inequality cannot produce a tighter bound even if we have more data (e.g., median).
Kolmogorov-Smirnov (KS) confidence limits\cite{saw1984chebyshev} also provide distribution-free bounds on an empirical CDF.
The calculation of KS limits requires, however, requires access to sample data.
\subsection{Limitations}
Our study has limitations in the following context.
First, we assume independence among the model parameters.
To the extent of our knowledge, how to model dependencies among the parameters in the context of uncertainty propagation using PBA and black-box models is an open problem.
One potential approach for modeling dependencies among parameters is to use a copula to represent the joint uncertainty of all parameters.\cite{joe1997multivariate}
A copula approach factors the joint CDF into a product of independent marginal CDFs and a copula that capture the dependencies.
In this formulation of bounds using a joint CDF, the overall bound is a function of the bounds on CDFs for some parameters represented by their p-boxes and the bounds of the copula.
The potentially promising approach using copula warrants further study and is, however, beyond the scope of our study.
Secondly, our study does not address the question of when one should consider using p-box vs. assuming a particular CDF to characterize uncertainty.
Instead of being prescriptive, we defer such decisions to the analysts because the level of uncertainty at which a p-box is the preferred approach is problem-dependent.
For example, a parameter may be highly uncertain due to the lack of empirical data and/or previous knowledge and, at the same time, non-influential, i.e., the model outcome is not sensitive to variations in the parameter values.
Thirdly, we provide a rudimentary treatment on how to make decisions using the results of a PBA.  
In situations where best-case/worst-case results are the basis for decision making, the analytical interval approach is preferred to assuming a distribution (e.g., uniform) and performing a simulation, particularly when that distribution may not correctly describe the parameters.
A more comprehensive treatment of decision-making based on interval values or bounds on probability distributions is needed, and it should be a focus of future studies on uncertainty quantification in decision-analytic modeling and cost-effectiveness analysis.
\section{Concluding remarks}
This study addresses limitations in current methodologies for characterizing uncertainty in data and knowledge used to inform mathematical models.
The novel methodology maximizes the use of existing limited information with the fewest number of assumptions, and provides a way to honestly characterize the uncertainty in the model parameters distributions used in decision-analytic modeling and cost-effectiveness analysis studies.

\newpage

\section*{acknowledgements}
I would like to thank Shaun Forbes and Thomas Trikalinos for introducing me to the world of imprecise probabilities or Knightian uncertainty. I would like to acknowledge the COMED consortium for their excellent support.
I am so thankful to Cassandra Berns for her help in deriving some of the inverses and plotting the figures.
I am grateful to Kosta Shatrov and Vishahan Suntharam for their reviews.
\section*{conflict of interest}
We have no conflict of interest to declare. 
\section*{Funding information}
This project received funding from the European Union's Horizon 2020 research and innovation programme under grant agreement $779306$ (COMED-Pushing the Boundaries of Cost and Outcome Analysis of Medical Technologies).




\bibliographystyle{unsrt}  
\bibliography{sample.bib}

\begin{thebibliography}{10}

\bibitem{hunink2014decision}
MG~Myriam Hunink, Milton~C Weinstein, Eve Wittenberg, Michael~F Drummond,
  Joseph~S Pliskin, John~B Wong, and Paul~P Glasziou.
\newblock {\em Decision making in health and medicine: integrating evidence and
  values}.
\newblock Cambridge University Press, 2014.

\bibitem{drummond2020modeling}
Michael~F Drummond.
\newblock Modeling in early stages of technology development: Is an iterative
  approach needed?: Comment on" problems and promises of health technologies:
  The role of early health economic modeling".
\newblock {\em International Journal of Health Policy and Management},
  9(6):260, 2020.

\bibitem{rothery2017characterising}
Claire Rothery, Karl Claxton, Stephen Palmer, David Epstein, Rosanna Tarricone,
  and Mark Sculpher.
\newblock Characterising uncertainty in the assessment of medical devices and
  determining future research needs.
\newblock {\em Health economics}, 26:109--123, 2017.

\bibitem{world2019guide}
World~Health Organization et~al.
\newblock Who guide for standardization of economic evaluations of immunization
  programmes.
\newblock Technical report, World Health Organization, 2019.

\bibitem{briggs2012model}
Andrew~H Briggs, Milton~C Weinstein, Elisabeth~AL Fenwick, Jonathan Karnon,
  Mark~J Sculpher, and A~David Paltiel.
\newblock Model parameter estimation and uncertainty analysis: a report of the
  ispor-smdm modeling good research practices task force working group--6.
\newblock {\em Medical decision making}, 32(5):722--732, 2012.

\bibitem{dahabreh2016recommendations}
Issa~J Dahabreh, Thomas~A Trikalinos, Ethan~M Balk, and John~B Wong.
\newblock Recommendations for the conduct and reporting of modeling and
  simulation studies in health technology assessment.
\newblock {\em Annals of internal medicine}, 165(8):575--581, 2016.

\bibitem{sanders2016recommendations}
Gillian~D Sanders, Peter~J Neumann, Anirban Basu, Dan~W Brock, David Feeny,
  Murray Krahn, Karen~M Kuntz, David~O Meltzer, Douglas~K Owens, Lisa~A
  Prosser, et~al.
\newblock Recommendations for conduct, methodological practices, and reporting
  of cost-effectiveness analyses: second panel on cost-effectiveness in health
  and medicine.
\newblock {\em Jama}, 316(10):1093--1103, 2016.

\bibitem{national2012assessing}
National~Research Council et~al.
\newblock {\em Assessing the reliability of complex models: mathematical and
  statistical foundations of verification, validation, and uncertainty
  quantification}.
\newblock National Academies Press, 2012.

\bibitem{helton2011quantification}
Jon~C Helton and Jay~D Johnson.
\newblock Quantification of margins and uncertainties: Alternative
  representations of epistemic uncertainty.
\newblock {\em Reliability Engineering \& System Safety}, 96(9):1034--1052,
  2011.

\bibitem{lee2009comparative}
Sang~Hoon Lee and Wei Chen.
\newblock A comparative study of uncertainty propagation methods for
  black-box-type problems.
\newblock {\em Structural and Multidisciplinary Optimization}, 37(3):239, 2009.

\bibitem{ohagan2006uncertain}
Anthony O'Hagan, Caitlin~E Buck, Alireza Daneshkhah, J~Richard Eiser, Paul~H
  Garthwaite, David~J Jenkinson, Jeremy~E Oakley, and Tim Rakow.
\newblock {\em Uncertain judgements: eliciting experts' probabilities}.
\newblock John Wiley \& Sons, 2006.

\bibitem{ferson2015constructing}
Scott Ferson, Vladik Kreinovich, Lev Grinzburg, Davis Myers, and Kari Sentz.
\newblock Constructing probability boxes and dempster-shafer structures.
\newblock Technical report, Sandia National Lab.(SNL-NM), Albuquerque, NM
  (United States), 2015.

\bibitem{liu2017efficient}
Xin Liu, Lairong Yin, Lin Hu, and Zhiyong Zhang.
\newblock An efficient reliability analysis approach for structure based on
  probability and probability box models.
\newblock {\em Structural and Multidisciplinary Optimization}, 56(1):167--181,
  2017.

\bibitem{beer2013imprecise}
Michael Beer, Scott Ferson, and Vladik Kreinovich.
\newblock Imprecise probabilities in engineering analyses.
\newblock {\em Mechanical systems and signal processing}, 37(1-2):4--29, 2013.

\bibitem{enszer2011probability}
Joshua~A Enszer, Youdong Lin, Scott Ferson, George~F Corliss, and Mark~A
  Stadtherr.
\newblock Probability bounds analysis for nonlinear dynamic process models.
\newblock {\em AIChE journal}, 57(2):404--422, 2011.

\bibitem{kriegler2005utilizing}
Elmar Kriegler and Hermann Held.
\newblock Utilizing belief functions for the estimation of future climate
  change.
\newblock {\em International journal of approximate reasoning},
  39(2-3):185--209, 2005.

\bibitem{nong2007estimation}
Andy Nong and Kannan Krishnan.
\newblock Estimation of interindividual pharmacokinetic variability factor for
  inhaled volatile organic chemicals using a probability-bounds approach.
\newblock {\em Regulatory Toxicology and Pharmacology}, 48(1):93--101, 2007.

\bibitem{iskandar2018theoretical}
Rowan Iskandar.
\newblock A theoretical foundation for state-transition cohort models in health
  decision analysis.
\newblock {\em PloS one}, 13(12):e0205543, 2018.

\bibitem{doubilet1985probabilistic}
Peter Doubilet, Colin~B Begg, Milton~C Weinstein, Peter Braun, and Barbara~J
  McNeil.
\newblock Probabilistic sensitivity analysis using monte carlo simulation: a
  practical approach.
\newblock {\em Medical decision making}, 5(2):157--177, 1985.

\bibitem{williamson1990probabilistic}
Robert~C Williamson and Tom Downs.
\newblock Probabilistic arithmetic. i. numerical methods for calculating
  convolutions and dependency bounds.
\newblock {\em International journal of approximate reasoning}, 4(2):89--158,
  1990.

\bibitem{ferson1996whereof}
Scott Ferson, Lev Ginzburg, and Resit Ak{\c{c}}akaya.
\newblock Whereof one cannot speak: when input distributions are unknown.
\newblock {\em Risk Analysis}, 1996.

\bibitem{zhang2010interval}
Hao Zhang, Robert~L Mullen, and Rafi~L Muhanna.
\newblock Interval monte carlo methods for structural reliability.
\newblock {\em Structural Safety}, 32(3):183--190, 2010.

\bibitem{rudin1964principles}
Walter Rudin et~al.
\newblock {\em Principles of mathematical analysis}, volume~3.
\newblock McGraw-hill New York, 1964.

\bibitem{cullen1999probabilistic}
Alison~C Cullen, H~Christopher Frey, and Christopher~H Frey.
\newblock {\em Probabilistic techniques in exposure assessment: a handbook for
  dealing with variability and uncertainty in models and inputs}.
\newblock Springer Science \& Business Media, 1999.

\bibitem{gilboa2009theory}
Itzhak Gilboa.
\newblock {\em Theory of decision under uncertainty}, volume~45.
\newblock Cambridge university press, 2009.

\bibitem{dong2006early}
Hengjin Dong and Martin Buxton.
\newblock Early assessment of the likely cost-effectiveness of a new
  technology: a markov model with probabilistic sensitivity analysis of
  computer-assisted total knee replacement.
\newblock {\em International Journal of Technology Assessment in Health Care},
  22(2):191--202, 2006.

\bibitem{R2019}
{R Core Team}.
\newblock {\em R: A Language and Environment for Statistical Computing}.
\newblock R Foundation for Statistical Computing, Vienna, Austria, 2019.

\bibitem{mak2018support}
Simon Mak, V~Roshan Joseph, et~al.
\newblock Support points.
\newblock {\em The Annals of Statistics}, 46(6A):2562--2592, 2018.

\bibitem{gablonsky2001locally}
Joerg~M Gablonsky and Carl~T Kelley.
\newblock A locally-biased form of the direct algorithm.
\newblock {\em Journal of Global Optimization}, 21(1):27--37, 2001.

\bibitem{johnson2014nlopt}
Steven~G Johnson.
\newblock The nlopt nonlinear-optimization package, 2014.

\bibitem{ypma2014introduction}
Jelmer Ypma.
\newblock Introduction to nloptr: an r interface to nlopt.
\newblock {\em R Package}, 2, 2014.

\bibitem{den2019guidelines}
Saskia Den~Boon, Mark Jit, Marc Brisson, Graham Medley, Philippe Beutels,
  Richard White, Stefan Flasche, T~D{\'e}irdre Hollingsworth, Tini Garske,
  Virginia~E Pitzer, et~al.
\newblock Guidelines for multi-model comparisons of the impact of infectious
  disease interventions.
\newblock {\em BMC medicine}, 17(1):163, 2019.

\bibitem{ferson2004arithmetic}
Scott Ferson and Janos~G Hajagos.
\newblock Arithmetic with uncertain numbers: rigorous and (often) best possible
  answers.
\newblock {\em Reliability Engineering \& System Safety}, 85(1-3):135--152,
  2004.

\bibitem{aughenbaugh2007probability}
Jason~Matthew Aughenbaugh and Christiaan~JJ Paredis.
\newblock Probability bounds analysis as a general approach to sensitivity
  analysis in decision making under uncertainty.
\newblock {\em SAE Transactions}, pages 1325--1339, 2007.

\bibitem{alefeld2012introduction}
Gotz Alefeld and Jurgen Herzberger.
\newblock {\em Introduction to interval computation}.
\newblock Academic press, 2012.

\bibitem{savage1972foundations}
Leonard~J Savage.
\newblock {\em The foundations of statistics}.
\newblock Courier Corporation, 1972.

\bibitem{tucker2003probability}
W~Troy Tucker and Scott Ferson.
\newblock Probability bounds analysis in environmental risk assessment.
\newblock {\em Applied Biomathematics, Setauket, New York}, 2003.

\bibitem{taleb2007black}
Nassim~Nicholas Taleb.
\newblock Black swans and the domains of statistics.
\newblock {\em The American Statistician}, 61(3):198--200, 2007.

\bibitem{norton2008ignorance}
John~D Norton.
\newblock Ignorance and indifference.
\newblock {\em Philosophy of Science}, 75(1):45--68, 2008.

\bibitem{laplace1812marquis}
Pierre-Simon Laplace.
\newblock Marquis de: Th{\'e}orie analytique des probabilit{\'e}s.
\newblock {\em Par M. le comte Laplace. Paris}, 1812.

\bibitem{schobi2017uncertainty}
Roland Sch{\"o}bi and Bruno Sudret.
\newblock Uncertainty propagation of p-boxes using sparse polynomial chaos
  expansions.
\newblock {\em Journal of Computational Physics}, 339:307--327, 2017.

\bibitem{deng2013improved}
Wu~Deng, Xinhua Yang, Li~Zou, Meng Wang, Yaqing Liu, and Yuanyuan Li.
\newblock An improved self-adaptive differential evolution algorithm and its
  application.
\newblock {\em Chemometrics and intelligent laboratory systems}, 128:66--76,
  2013.

\bibitem{ellis2020active}
Alexandra~G Ellis, Rowan Iskandar, Christopher~H Schmid, John~B Wong, and
  Thomas~A Trikalinos.
\newblock Active learning for efficiently training emulators of computationally
  expensive mathematical models.
\newblock {\em Statistics in Medicine}, 39(25):3521--3548, 2020.

\bibitem{augustin2014introduction}
Thomas Augustin, Frank~PA Coolen, Gert De~Cooman, and Matthias~CM Troffaes.
\newblock {\em Introduction to imprecise probabilities}.
\newblock John Wiley \& Sons, 2014.

\bibitem{dempster2008upper}
Arthur~P Dempster.
\newblock Upper and lower probabilities induced by a multivalued mapping.
\newblock In {\em Classic works of the Dempster-Shafer theory of belief
  functions}, pages 57--72. Springer, 2008.

\bibitem{shafer1976mathematical}
Glenn Shafer.
\newblock {\em A mathematical theory of evidence}, volume~42.
\newblock Princeton university press, 1976.

\bibitem{berger1990robust}
James~O Berger.
\newblock Robust bayesian analysis: sensitivity to the prior.
\newblock {\em Journal of statistical planning and inference}, 25(3):303--328,
  1990.

\bibitem{saw1984chebyshev}
John~G Saw, Mark~CK Yang, and Tse~Chin Mo.
\newblock Chebyshev inequality with estimated mean and variance.
\newblock {\em The American Statistician}, 38(2):130--132, 1984.

\bibitem{joe1997multivariate}
Harry Joe.
\newblock {\em Multivariate models and multivariate dependence concepts}.
\newblock CRC Press, 1997.

\end{thebibliography}
\appendix
\section{Appendices}
\renewcommand{\thefigure}{\thesubsection.\arabic{figure}}
\numberwithin{equation}{section}
\setcounter{equation}{0}
\setcounter{figure}{0} 

\subsection{Derivation of p-box formulas}\label{app:pbox_derivation}
In this section, we show a heuristic approach for deriving a p-box for $\mathcal{D}=\{\ltt,\utt,\mu,\sigma\}$ as an exemplar of other $\mathcal{D}$.
We use the same notations and definitions as described in the main text.
\\ \\
The LBF of the p-box,$\mathcal{P}_{\ltt,\utt,\mu,\sigma}$ is given by:
\begin{align}
    \lF = \begin{cases}
     0 \quad &\mbox{ for } \quad \theta < \xi_1 \\
    \frac{\sigma^2+(b-\mu)(\theta-\mu)}{(\utt-\ltt)(\theta-\ltt)} \quad &\mbox{ for } \quad \xi_1 \le \theta <\xi_2\\
        \frac{(\theta-\mu)^2}{(\theta-\mu)^2 + \sigma^2} \quad &\mbox{ for } \quad \xi_2 \le \theta < \utt\\
    1 \quad &\mbox{ for } \quad b \le \theta\\
    \end{cases}
\end{align}
Let $\mc{F}$ denote the set of \textit{all} CDFs ($\Ft$), that are consistent with the minimal data: $\{\ltt,\utt,\mu,\sigma\}$.
From the definitions of a CDF, any $\Ft \in \mc{F}$ satisfies: $0 \le \Ft \le 1$, $F(\lt)=0$ , and $F(\ut)=1$ (denoted collectively as Assumption (*)).
Let $ft$ be the probability density function.
Using the formula for a mean, we obtain the following equation:
\begin{align*}
    \mu = b F(b) - a F(a) - \intt \Ftt d\theta = b - \intt \Ftt d\theta
\end{align*}
and the following relation:
\begin{align} \label{eq:app_S}
    \intt\Ftt d\theta = b -\mu = S(\theta)
\end{align}
Since we know the standard deviation $\sigma$, we have:
\begin{align*}
    \sigma^2&=\intt \theta^2 \ftt d\theta - \mu^2 \\
    &=2 \intt S(\theta) d\theta - (\utt-\mu)^2
\end{align*}
or
\begin{align} \label{eq:app_std}
    \intt S(\theta) d\theta = \frac{1}{2}\left(\sigma^2+(b-\mu)^2 \right)
\end{align}
Since the derivative of $S(\theta)$ is equal $\Ft$, then:
\begin{align} \label{eq:app_Sprime}
    0 \le S'(\theta) \le 1
\end{align}
To derive the LBF, we consider three cases depending on the value of $\theta$.
The first case is $a \le \theta \le \xi_1$.
In this interval, $\Ft$ has a minimum of $0$.
We want to (1) find $\Ft \in \mc{F}$ that (i) satisfies Equations \ref{eq:app_S}, \ref{eq:app_std},and \ref{eq:app_Sprime} as well as (*) and (ii) takes the value of $0$ $\forall \theta \le \xi_1$ and (2) determine the maximum value of $\xi_1$.
For $\xi_1$ to be a maximum, the following has to be true: the maximum area under $S(\theta)$ is concentrated in $\xi_1 < \theta \le b$.
Since $S'(\theta)=\Ft$ is non-decreasing, the maximum area will be obtained if $\Ft$ is constant (or $S(\theta)$ is linear) over $\xi_1 < \theta \le b$.
Using Equation \ref{eq:app_std} to compute the area under $S(\theta)$, we obtain the following equality:
\begin{align*}
    \frac{1}{2}(b-\xi_1)(b-\mu)=\frac{1}{2}\left(\sigma^2+(b-\mu)^2\right)
\end{align*}
Solving for $\xi_1$, we have:
\begin{align} \label{eq:app_xi1}
    \xi_1=\mu-\frac{\sigma^2}{b-\mu}
\end{align}
Therefore, for $x \le \xi_1$ and $x > \xi_1$, $\lF=0$ and $\lF=\frac{b-\mu}{b-\xi_1}=\frac{(b-\mu)^2}{(b-\mu)^2+\sigma^2}$, respectively.
\\ \\
The second case is $\xi_1 < \theta \le \xi_2$.
Since $\theta > \xi_1$, then $\Ft \neq 0$.
To find the minimum possible $\Ft$, we need to concentrate the maximum area under $\Ft$ to the left of $\theta$.
This is only possible when $\Ft$ is constant (similar to the above) and takes different (constant) values, depending on whether the interval is to the left or right of $\theta$.
Geometrically, the shape of $\Ftt$ follows a step function.
Using Equation \ref{eq:app_std} to compute the area under $S(\theta)$, we obtain the following equality:
\begin{align*}
    \frac{1}{2}(\theta-a)S(\theta)+\frac{1}{2}(S(\theta)+b-\mu)(b-\theta)=\frac{1}{2}\left(\sigma^2+(b-\mu)^2\right).
\end{align*}
Hence, we have:
\begin{align*}
    \Ftt=\frac{S(\theta)}{\theta-a}=\frac{\sigma^2+(b-\mu)(\theta-\mu)}{(b-a)(\theta-a)}.
\end{align*}
Thus, for $a < \theta' \le \theta$:
\begin{align} \label{eq:app_Fa}
    F(\theta')=\frac{\sigma^2+(b-\mu)(\theta-\mu)}{(b-a)(\theta-a)}
\end{align}
and, for $\theta < \theta' <b$, using Equation \ref{eq:app_S}, we have:
\begin{align} \label{eq:app_Fb}
        F(\theta')=\frac{(b-\mu)(b-a+\mu-\theta)-\sigma^2}{(b-a)(b-\theta)}.
\end{align}
Since $\Ftt$ must be non-decreasing, Equation \ref{eq:app_Fb} must be no less than Equation \ref{eq:app_Fa}:
\begin{align} \label{eq:app_cond1}
    (b-\mu)(\mu-a) \ge \sigma^2;
\end{align}
otherwise, Equations \ref{eq:app_S}, \ref{eq:app_std},and \ref{eq:app_Sprime} as well as (*) are not satisfied.
We also have another condition, i.e., Equation \ref{eq:app_Fb} must not exceed 1, or:
\begin{align} \label{eq:}
    \theta \le \mu+\frac{\sigma^2}{\mu-a}=\xi_2.
\end{align}
In sum, for $\xi_1 < \theta \le \xi_2$, we obtain:
\begin{align}
    \lF = \frac{\sigma^2+(b-\mu)(\theta-\mu)}{(b-a)(\theta-a)}
\end{align}
\\ \\
The third case is $\xi_2 < \theta \le b$.
Since $\theta > \xi_2$, $F(\theta')$ must be equal to 1 for $\theta'>\theta$.
In contrast to the second case, for $\theta'$ to the left of $\theta$, $F(\theta')$ can not be strictly constant.
To obtain the minimum value of $F(\theta')$, we split the interval into $\theta' \le \xi_0$ and $\theta'> \xi_0$ for some $\xi_0$: $F(\theta')=0$ for $\theta' \le \xi_0$ and  $ F(\theta')$ is constant for $\theta' \le \xi_0$. 
As above, using Equation \ref{eq:app_std} to compute the area under $S(\theta)$, we obtain the following equality:
\begin{align*}
    \frac{1}{2}(\theta-\xi_0)S(\theta)+\frac{1}{2}(S(\theta)+b-\mu)
    )(b-\theta)=\frac{1}{2}(\sigma^2+(b-\mu)^2),
\end{align*}
or
\begin{align*}
    \xi_0 = \mu - \frac{\sigma^2}{\theta-\mu}
\end{align*}
Therefore, the minimum value of $\Ftt$ is given by:
\begin{align*}
    F(\theta)=\frac{(\theta-\mu)^2}{(\theta-\mu)^2+\sigma^2} = \lF
\end{align*}
Combining the results of the three cases, we obtain Equation \ref{eq:pbox_std_LBF}.
The derivation of the UBF follows identical steps: instead of a minimum, we find the maximum of $\Ft$ and start the derivation from the maximum value of $\theta$ ($b$) . The derivations of p-boxes for other $\mc{D}$ follow the same principles, i.e., by ensuring that certain constraints are respected.

\subsection{P-box as the tightest bound}\label{app:pbox_tight}
In this section,we show that a p-box gives the tightest bounds on the unknown CDF, given $\mathcal{D}=\{a,b,m\}$ as an exemplar of other $\mathcal{D}$.
We consider only the LBF as the procedure for the UBF follows the same steps. The LBF for $\mathcal{D}=\{a,b,m\}$ is given by:
\begin{align*}
    \lF = \begin{cases}
     0 \quad &\mbox{ for } \quad \theta < m \\
    0.5 \quad &\mbox{ for } \quad m \le \theta < \ut\\
        1 \quad &\mbox{ for } \quad \ut \le \theta\\
    \end{cases}
\end{align*}
Let $\mc{F}$ denote the set of \textit{all} CDFs ($\Ft$), that have $\{a,b,m\}$.
From the definitions of a CDF, any $\Ft \in \mc{F}$ satisfies: $0 \le \Ft \le 1$, $F(\lt)=0$ , and $F(\ut)=1$ (denoted collectively as Assumption (*)).
We use a proof by contradiction.
Suppose that there exist an LBF $\lG$ (different from $\lF$) and a UBF $\uG$ (different from $\uF$) such that the following is true:
\begin{align}\label{eq:app_equality}
    \lF\le \lG \le \Ftt \le \uG \uF
\end{align}
Consequently, there exist $\theta_1 \in [a,b]$ such that the following holds true:
\begin{align} \label{eq:app_equality2}
     \underbar{F}(\theta_1) < \underbar{G}(\theta_1) \le F(\theta_1)
\end{align}
We consider two cases: $\theta<m$ and $\theta \ge m$.
From Equation \ref{eq: pbox_median_LBF}, we know $\lF=0$ for $\theta <m$.
Therefore, we can find an $\theta_1\in [a,m)$ such that: 
\begin{align}\label{eq:app_equality3}
0<\underbar{G}(\theta_1) \le \underbar{F}(x_1)
\end{align}
We pick an $F(\theta) \in \mc{F}$ with the following form:
\begin{align}\label{eq:app_F1}
    F(\theta)=\begin{cases}
     \frac{\underbar{G}(\theta_1)}{2(\theta_1-a)(\theta-a)}&\mbox{ for } \quad a \le \theta <\theta_1\\
    \frac{1}{2}\underbar{G}(\theta_1)+\frac{1-\underbar{G}(\theta_1)}{2(m-\theta_1)}(\theta-\theta_1) &\mbox{ for } \quad \theta_1 \le \theta <m
    \end{cases}
\end{align}
The form is chosen specifically for ensuring $F(\theta_1)=\frac{1}{2}\underbar{G}(\theta_1)$.
Therefore, we have $\underbar{G}(\theta_1)>\underbar{F}(\theta_1)$ which contradicts Equation \ref{eq:app_equality3}.
In sum, we are able to find $F(\theta)\in\mc{F}$ that does not satisfy 
the inequality in Equation \ref{eq:app_equality} for the proposed lower bound $\underbar{G}(\theta)$ in the interval $\theta <m$.
Furthermore, Equation \ref{eq:app_equality} is violated for any value of $\underbar{G}(\theta)>0$.
Hence, the lower boundary for $\theta<m$ can not take positive values which contradicts our assumption about the existence of $\underbar{G}(\theta)$ that is different than $\lF$.
\\ \\
We follow the same approach for $\theta \ge m$.
From Equation \ref{eq: pbox_median_LBF},$\lF=\frac{1}{2}$.
Therefore, we can find an $\theta_1\in [m,b]$ such that: 
\begin{align}\label{eq:app_equality4}
\frac{1}{2}<\underbar{G}(\theta_1) \le \underbar{F}(x_1)
\end{align}  
We choose an $F(\theta) \in \mc{F}$ with the following form:
\begin{align}\label{eq:app_F1}
    F(\theta)=\begin{cases}
     \frac{1}{2}+\frac{\underbar{G}(\theta_1)-\frac{1}{2}}{2(\theta_1-m)(\theta-m)}&\mbox{ for } \quad m \le \theta <\theta_1\\
    \frac{1}{4}+\frac{1}{2}\underbar{G}(\theta_1)+\frac{\frac{3}{2}-\underbar{G}(\theta_1)}{2(b-\theta_1)}(\theta-\theta_1) &\mbox{ for } \quad \theta_1 \le \theta \le b
    \end{cases}
\end{align}
The form is chosen specifically for ensuring $F(\theta_1)=\frac{1}{2}+\frac{1}{2}\left(\underbar{G}(\theta_1)-\frac{1}{2}\right)$.
As before, for the proposed lower bound $\underbar{G}(\theta)$, we can find an $\lFF \in \mc{F}$ that violates inequality (Equation\ref{eq:app_equality}.
Furthermore, Equation \ref{eq:app_equality} cannot be satisfied for any value of $\underbar{G}(\theta)>\frac{1}{2}$.
Thus, the lower boundary of $\theta \ge m$ cannot take any values $> \frac{1}{2}$; thereby contradicting our assumption about the existence of $\underbar{G}(\theta)$ that is different than $\lF$.

\subsection{P-box formula for minimum, maximum, mean, and median case}\label{app:pbox_median_mean}
This sections shows the formula for a p-box for $\mathcal{D}=\{\ltt,\utt,m, \mu\}$.
We use the same notations and definitions as described in the main text.
\\ \\
We define the following new notations:
\begin{align*}
    c&=\frac{a+b}{2} \\
    \phi&=\frac{b-\mu}{b-a} \\
    \gamma &= 2 \mu - a \\
    \psi &= 2 \mu - b \\
    \mu_{min}&=\frac{a+m}{2} \\
    \mu_{min}&=\frac{m+b}{2} \\
\end{align*}
The formulas of the p-box depend on the values of the above variables:
\begin{enumerate}
    \item $m<\mu$, $\mu < c$, $\phi > \frac{1}{2}$
\begin{align} \label{eq:pbox_median_mean_1}
    \lF = \begin{cases}
     0 \quad &\mbox{ for } \quad \theta < m \\
    \frac{1}{2} \quad &\mbox{ for } \quad m \le \theta < \gamma\\
    \frac{\theta-\mu}{\theta-a} \quad &\mbox{ for } \quad \theta \ge \gamma\\
    \end{cases},
\quad \quad
    \uF = \begin{cases}
    \frac{1}{2}  \quad &\mbox{ for } \quad \theta \le m \\
    \frac{b-\mu}{b-\theta} \quad &\mbox{ for } \quad m < \theta < \mu\\
    1 \quad &\mbox{ for } \quad \theta \ge \mu\\
    \end{cases}
\end{align}
    \item $m<\mu$, $\mu=c$,$\phi = \frac{1}{2}$
\begin{align} \label{eq:pbox_median_mean_2}
    \lF = \begin{cases}
     0 \quad &\mbox{ for } \quad \theta < m \\
    \frac{1}{2} \quad &\mbox{ for } \quad \theta \ge m \\
    \end{cases},
\quad \quad
    \uF = \begin{cases}
    \frac{1}{2}  \quad &\mbox{ for } \quad \theta \le m \\
    \frac{b-\mu}{b-\theta} \quad &\mbox{ for } \quad m < \theta < \mu\\
    1 \quad &\mbox{ for } \quad \theta \ge \mu\\
    \end{cases}
\end{align}
    \item $m<\mu$, $c <\mu < \mu_{max}$, $\phi < \frac{1}{2}$, $m>\psi$
\begin{align} \label{eq:pbox_median_mean_3}
    \lF = \begin{cases}
     0 \quad &\mbox{ for } \quad \theta < m \\
    \frac{1}{2} \quad &\mbox{ for } \quad \theta \ge m \\
    \end{cases},
\quad \quad
    \uF = \begin{cases}
    \frac{b-\mu}{b-\theta}  \quad &\mbox{ for } \quad \theta \le \psi \mbox{ or }  m < \theta < \mu \\
    \frac{1}{2} \quad &\mbox{ for } \quad \psi < \theta \le m\\
    1 \quad &\mbox{ for } \quad \theta \ge \mu\\
    \end{cases}
\end{align}
    \item $m<\mu$, $\mu=\mu_{max}$, $\phi<\frac{1}{2}$, $m=\psi$
\begin{align} \label{eq:pbox_median_mean_4}
    \lF = \begin{cases}
     0 \quad &\mbox{ for } \quad \theta < m \\
    \frac{1}{2} \quad &\mbox{ for } \quad \theta \ge m \\
    \end{cases},
\quad \quad
    \uF = \begin{cases}
    \frac{b-\mu}{b-\theta}  \quad &\mbox{ for } \quad  \theta < \mu \\
    1 \quad &\mbox{ for } \quad \theta \ge \mu\\
    \end{cases}
\end{align}
    \item $m=\mu$, $\mu < c$, $\phi > \frac{1}{2}$
\begin{align} \label{eq:pbox_median_mean_5}
    \lF = \begin{cases}
    0 \quad &\mbox{ for } \quad \theta \le \psi \mbox{ or }  m < \theta < \mu \\
    \frac{1}{2} \quad &\mbox{ for } \quad \psi < \theta \le m\\
    \frac{\theta-\mu}{\theta-a}  \quad &\mbox{ for } \quad \theta \ge \phi\\
    \end{cases},
\quad \quad
    \uF = \begin{cases}
    \frac{1}{2} \quad &\mbox{ for } \quad  \theta \le m\\
    1 \quad &\mbox{ for } \quad \theta > m \\
    \end{cases}
\end{align}
    \item $m=\mu$, $\mu=c$,  $\phi = \frac{1}{2}$
\begin{align} \label{eq:pbox_median_mean_6}
    \lF = \begin{cases}
    0 \quad &\mbox{ for } \quad \theta < m \\
    \frac{1}{2} \quad &\mbox{ for } \quad \theta \ge m\\
    \end{cases},
\quad \quad
    \uF = \begin{cases}
    \frac{1}{2} \quad &\mbox{ for } \quad  \theta \le m\\
    1 \quad &\mbox{ for } \quad \theta > m \\
    \end{cases}
\end{align}
    \item $m=\mu$, $\mu>c$, $\phi < \frac{1}{2}$
\begin{align} \label{eq:pbox_median_mean_7}
    \lF = \begin{cases}
    0 \quad &\mbox{ for } \quad \theta < m \\
    \frac{1}{2} \quad &\mbox{ for } \quad \theta \ge m\\
    \end{cases},
\quad \quad
    \uF = \begin{cases}
    \frac{b-\mu}{b-\theta}  \quad &\mbox{ for } \quad  \theta \le \psi \\
    \frac{1}{2} \quad &\mbox{ for } \quad  \psi < \theta \le m\\
    1 \quad &\mbox{ for } \quad \theta > m \\
    \end{cases}
\end{align}
    \item $m > \mu$, $\mu=\mu_{min}$,$\phi > \frac{1}{2}$, $m=\psi$
\begin{align} \label{eq:pbox_median_mean_8}
    \lF = \begin{cases}
    0 \quad &\mbox{ for } \quad \theta < \mu \\
    \frac{\theta-\mu}{\theta-a}  \quad &\mbox{ for } \quad \theta \ge \mu\\
    \end{cases},
\quad \quad
    \uF = \begin{cases}
    \frac{1}{2} \quad &\mbox{ for } \quad  \psi < \theta \le m\\
    1 \quad &\mbox{ for } \quad \theta > m \\
    \end{cases}
\end{align}
    \item $m > \mu$, $\mu_{min}<\mu<c$,$\phi > \frac{1}{2}$, $m<\psi$
\begin{align} \label{eq:pbox_median_mean_9}
    \lF = \begin{cases}
    0 \quad &\mbox{ for } \quad \theta < \mu \\
    \frac{\theta-\mu}{\theta-a}  \quad &\mbox{ for } \quad \mu \le \theta < m \mbox{ or } \theta \ge \phi\\
        \frac{1}{2} \quad &\mbox{ for } \quad m \le \theta < \phi  \\
    \end{cases},
\quad \quad
    \uF = \begin{cases}
    \frac{1}{2} \quad &\mbox{ for } \quad  \psi < \theta \le m\\
    1 \quad &\mbox{ for } \quad \theta > m \\
    \end{cases}
\end{align}
    \item $m > \mu$, $\mu=c$,$\phi = \frac{1}{2}$
\begin{align} \label{eq:pbox_median_mean_10}
    \lF = \begin{cases}
    0 \quad &\mbox{ for } \quad \theta < \mu \\
    \frac{\theta-\mu}{\theta-a}  \quad &\mbox{ for } \quad \mu \le \theta < m \\
    \frac{1}{2} \quad &\mbox{ for } \quad \theta \ge m \\
    \end{cases},
\quad \quad
    \uF = \begin{cases}
    \frac{1}{2} \quad &\mbox{ for } \quad  \psi < \theta \le m\\
    1 \quad &\mbox{ for } \quad \theta > m \\
    \end{cases}
\end{align}
    \item $m > \mu$, $\mu>c$,$\phi < \frac{1}{2}$
\begin{align} \label{eq:pbox_median_mean_11}
    \lF = \begin{cases}
    0 \quad &\mbox{ for } \quad \theta < \mu \\
    \frac{\theta-\mu}{\theta-a}  \quad &\mbox{ for } \quad \mu \le \theta < m \\
    \frac{1}{2} \quad &\mbox{ for } \quad \theta \ge m \\
    \end{cases},
\quad \quad
    \uF = \begin{cases}
    \frac{b-\mu}{b-\theta}  \quad &\mbox{ for } \quad \theta \le \psi \\
    \frac{1}{2} \quad &\mbox{ for } \quad  \psi < \theta \le m\\
    1 \quad &\mbox{ for } \quad \theta > m \\
    \end{cases}
\end{align}
\end{enumerate}

\subsection{Quasi-inverse of p-box formulas}\label{app:pbox_inverses}
The inverse functions of the p-box $\mathcal{P}_{\ltt,\utt}$ are given by:
\begin{align}
    \lFp^{-1}_{\ltt,\utt} = \begin{cases}
     [\ltt , \utt] \quad &\mbox{ for } \quad p = 0 \\
    \utt \quad &\mbox{ for } \quad 0 < p \le 1\\
    \end{cases}
\end{align}
for LBF, and,
\begin{align}
    \uFp^{-1}_{\ltt,\utt} = \begin{cases}
     \ltt \quad &\mbox{ for } \quad 0 \le p < 1 \\
    [\ltt, \utt] \quad &\mbox{ for } \quad p = 1\\
    \end{cases}
\end{align}
for UBF.
\\ \\
The inverse functions of the p-box $\mathcal{P}_{\ltt,\utt,m}$ are given by:
\begin{align}\label{eq:pbox_median_LBF_inv}
    \lFp^{-1}_{\ltt,\utt,m} = \begin{cases}
     [\ltt, m] \quad &\mbox{ for } \quad p = 0 \\
   m \quad &\mbox{ for } \quad 0 < p < 0.5\\
        [m, \utt] \quad &\mbox{ for } \quad p = 0.5 \\
   \utt \quad &\mbox{ for } \quad 0.5 < p \le 1\\
    \end{cases}
\end{align}
for LBF, and,
\begin{align}\label{eq:pbox_median_LBF_inv}
    \uFp^{-1}_{\ltt,\utt,m} = \begin{cases}
     \ltt \quad &\mbox{ for } \quad 0 \le p < 0.5 \\
   [\ltt, m] \quad &\mbox{ for } \quad p = 0.5\\
        m \quad &\mbox{ for } \quad 0.5 < p < 1\\
   [m, \utt] \quad &\mbox{ for } \quad p = 1\\
    \end{cases}
\end{align}
for UBF. \\ \\
The inverse functions of the p-box $\mathcal{P}_{\ltt,\utt,\mu}$ are given by:
\begin{align} \label{eq:pbox_mean_LBF_inv}
    \lFp^{-1}_{\ltt,\utt,\mu} = \begin{cases}
    [\ltt, \mu] \quad &\mbox{ for } \quad p = 0 \\
    \frac{p*\ltt-\mu}{p-1} \quad &\mbox{ for } \quad 0 < p < \frac{\utt - \mu}{\utt - \ltt}\\
    \utt \quad &\mbox{ for } \quad \frac{\utt - \mu}{\utt - \ltt} \le p \le 1\\
    \end{cases}
\end{align}
for LBF, and,
\begin{align} \label{eq:pbox_mean_LBF_inv}
    \uFp^{-1}_{\ltt,\utt,\mu} = \begin{cases}
    \ltt \quad &\mbox{ for } \quad 0 \le p \le \frac{\utt - \mu}{\utt - \ltt} \\
    \utt - \frac{\utt-\mu}{p} \quad &\mbox{ for } \quad \frac{\utt - \mu}{\utt - \ltt} < p < 1\\
    [\mu, \utt]\quad &\mbox{ for }\quad p = 1\\
    \end{cases}
\end{align}
for UBF. \\ \\
The inverse functions of the p-box $\mathcal{P}_{\ltt,\utt,\mu,\sigma}$ are given by:
\begin{align} \label{eq:pbox_std_LBF_inv}
    \lFp^{-1} = \begin{cases}
     [\ltt, \xi_1] \quad &\mbox{ for } \quad p=0 \\
    \frac{p\ltt(\utt-\ltt)-\mu(\utt-\mu)+\sigma^2}{p(\utt-\ltt)-(\utt-\mu)} \quad &\mbox{ for } \quad 0 < p \le  \frac{\sigma^2(\utt-\mu)\frac{\sigma^2}{\mu-\ltt}}{(\utt-\ltt)(\mu-\ltt+\frac{\sigma^2}{\mu-\ltt})}\\
        \mu + \sqrt{\frac{p\sigma^2}{1-p}} \quad &\mbox{ for } \quad \frac{\sigma^2(\utt-\mu)\frac{\sigma^2}{\mu-\ltt}}{(\utt-\ltt)(\mu-\ltt+\frac{\sigma^2}{\mu-\ltt})} < p < 1\\
    b \quad &\mbox{ for } \quad p=1\\
    \end{cases}
\end{align}
for LBF, and,
\begin{align} \label{eq:pbox_std_UBF_inv}
    \uFp^{-1} = \begin{cases}
         \ltt \quad &\mbox{ for } \quad p=0 \\
      \mu - \sqrt{\frac{\sigma^2(1-p)}{p}}  \quad &\mbox{ for } \quad 0<p \le \frac{\sigma^2}{(\frac{\sigma^2}{\utt-\mu})^2+\sigma^2} \\
    \frac{(\utt-\mu)(\utt-\ltt+\mu)-\sigma^2-p\utt(\utt-\ltt)}{(\utt-\mu)-p(\utt-\ltt)} \quad &\mbox{ for } \quad \frac{\sigma^2}{(\frac{\sigma^2}{\utt-\mu})^2+\sigma^2}<p<1\\
       [\xi_2, \utt] \quad &\mbox{ for } \quad p=1\\
    \end{cases}
\end{align}
for UBF, where $\xi_1 = \mu - \frac{\sigma^2}{b-\mu}$ and $\xi_2 = \mu + \frac{\sigma^2}{\mu-a}$.
\end{document}